%% file: vels.tex
\PassOptionsToPackage{unicode=true}{hyperref} 
\PassOptionsToPackage{hyphens}{url}
\documentclass[astrosymb, twocolumn, tighten]{aastex63}
\usepackage{lmodern}
\usepackage{amssymb,amsmath}
\usepackage{ifxetex,ifluatex}
\usepackage{fixltx2e} 
\ifnum 0\ifxetex 1\fi\ifluatex 1\fi=0 
  \usepackage[T1]{fontenc}
  \usepackage[utf8]{inputenc}
  \usepackage{textcomp} 
\else 
  \usepackage{unicode-math}
  \defaultfontfeatures{Ligatures=TeX,Scale=MatchLowercase}
\fi
\IfFileExists{upquote.sty}{\usepackage{upquote}}{}
\IfFileExists{microtype.sty}{%
\usepackage[]{microtype}
\UseMicrotypeSet[protrusion]{basicmath} 
}{}
\IfFileExists{parskip.sty}{%
\usepackage{parskip}
}{
\setlength{\parindent}{0pt}
\setlength{\parskip}{6pt plus 2pt minus 1pt}
}
\usepackage{hyperref}
\hypersetup{
            pdfborder={0 0 0},
            breaklinks=true}
\usepackage{listings}

\usepackage{graphicx,grffile}
\makeatletter
\def\maxwidth{\ifdim\Gin@nat@width>\linewidth\linewidth\else\Gin@nat@width\fi}
\def\maxheight{\ifdim\Gin@nat@height>\textheight\textheight\else\Gin@nat@height\fi}
\makeatother
\setkeys{Gin}{width=\maxwidth,height=\maxheight,keepaspectratio}
\providecommand{\tightlist}{%
  \setlength{\itemsep}{0pt}\setlength{\parskip}{0pt}}
\ifx\paragraph\undefined\else
\let\oldparagraph\paragraph
\renewcommand{\paragraph}[1]{\oldparagraph{#1}\mbox{}}
\fi
\ifx\subparagraph\undefined\else
\let\oldsubparagraph\subparagraph
\renewcommand{\subparagraph}[1]{\oldsubparagraph{#1}\mbox{}}
\fi

\makeatletter
\def\fps@figure{htbp}
\makeatother

\usepackage[capitalize]{cleveref}
\usepackage{CJKutf8}
\usepackage{booktabs}
\usepackage{mathptmx,txfonts,tikz}
\usepackage[outline]{contour}
\usetikzlibrary{shapes.geometric,arrows, calc,positioning, fit}
\definecolor{forestgreen}{rgb}{0.09, 0.44, 0.09}
\date{\today}

\begin{document}

\input{meta}

\begin{abstract}
The EXtreme PREcision Spectrograph (EXPRES) is an environmentally stabilized, fiber-fed, $R=137,500$, optical spectrograph. It was recently commissioned at the 4.3-m Lowell Discovery Telescope (LDT) near Flagstaff, Arizona. The spectrograph was designed with a target radial-velocity (RV) precision of 30\cms. In addition to instrumental innovations, the EXPRES pipeline, presented here, is the first for an on-sky, optical, fiber-fed spectrograph to employ many novel techniques---including an ``extended flat'' fiber used for wavelength-dependent quantum efficiency characterization of the CCD, a flat-relative optimal extraction algorithm, chromatic barycentric corrections, chromatic calibration offsets, and an ultra-precise laser frequency comb for wavelength calibration. We describe the reduction, calibration, and radial-velocity analysis pipeline used for EXPRES and present an example of our current sub-meter-per-second RV measurement precision, which reaches a formal, single-measurement error of 0.3\ms for an observation with a per-pixel signal-to-noise ratio of 250. These velocities yield an orbital solution on the known exoplanet host 51 Peg that matches literature values with a residual RMS of 0.895\ms.

\end{abstract}

\hypertarget{introduction}{%
\section{Introduction}\label{introduction}}

Results from the NASA \emph{Kepler} mission show that small planets with radii between 1--4 $R_{\oplus}$ are found orbiting 20--50\% of main-sequence stars \citep{winn_occurrence_2015}. While such transit surveys, such as \emph{Kepler}, K2, and TESS, have revealed a wealth of planets, few of these planets have had their masses measured; mass estimates which do exist are typically radial-velocity (RV), dynamical masses. More precise RV measurements are required to determine mass estimates for these planets, particularly small rocky ones, than are possible with pre-existing RV spectrographs. Were they available, these mass measurements would shed light on planetary structure, bulk density, and the mass-radius relation for sub-Neptune-mass planets. 

To meet these needs, a new generation of RV spectrographs is now emerging: the Echelle SPectrograph for Rocky Exoplanets Search and Stable Spectroscopic Observations \citep[ESPRESSO:][]{pepe_espresso_2013} and The EXtreme PREcision Spectrograph \citep[EXPRES:][]{jurgenson_expres_2016} are now on sky taking data, while the NN-explore Exoplanet Investigations with Doppler spectroscopy spectrograph \citep[NEID:][]{neid} is in the commissioning phase. These new extreme-precision radial velocity (EPRV) spectrographs are driving towards the sub-10-\cms{} radial-velocity precision needed to detect a true Earth twin around a Sun-like star.

Extremely precise spectrographs in turn demand extremely precise data reduction pipelines. The fidelity of the data and error estimates returned by these data reduction pipelines is paramount if groundbreaking discoveries returned from such instruments are to be credible. To this end, a complete science reduction, extraction, and analysis pipeline was newly developed and tailored for EXPRES data. In what follows, we begin with a brief description of the instrument and the calibration strategy. We then describe the analysis performed by our pipeline on EXPRES data, and present the first radial velocity measurements from the instrument.

\hypertarget{instrument-description}{%
\section{Instrument Description}\label{instrument-description}}

The EXtreme PREcision Spectrograph (EXPRES) is a fiber-fed, white-pupil EPRV spectrograph with a design resolution of $R = 150,000$ and a wavelength range of $3800-7800\unit{\AA}$. In practice, \citet{blackman_performance_2020} show that the median resolution is better characterized as $R \sim 137,500$ with a maximum resolution reaching $R \sim 150,000$ in some regions of the detector. EXPRES is environmentally stabilized in a vacuum enclosure and is situated at Lowell Observatory's 4.3-m Lowell Discovery Telescope (LDT) near Flagstaff, Arizona. The multi-instrument port configuration of the LDT allows for high-cadence, flexible scheduling of stars (up to 280 partial nights per year).

Owing to various changes to the hardware configuration during commissioning, we have divided radial velocities from EXPRES into different calibration epochs, with an independent RV offset fitted for each epoch. Each epoch demarcates changes introduced to either the configuration of the echellogram on the CCD, the shape of the PSF, or the stability of our calibration sources. A summary of the changes that delineate the beginning of each epoch can be found in \cref{tab:epochs}.

\begin{table}[ht!]
\scriptsize
\centering
\caption{Instrumental Epochs\label{tab:epochs}}
\begin{tabular}{cccl}
\toprule
Epoch & Start & End & Changes Before Epoch Start \tabularnewline
\midrule
0 & - & 2018-04-15 & Commissioning \tabularnewline
1 & 2018-04-15 & 2018-06-15 & Increased CCD pre-settle time \tabularnewline
2 & 2018-06-15 & 2018-11-08 & Fiber change; \tabularnewline
 & & & CCD rotation \tabularnewline
3 & 2018-11-08 & 2019-02-07 & Original science fiber replaced; \tabularnewline
 & & &  Calibration unit rebuilt \tabularnewline
4 & 2019-02-07 & 2019-08-04 & LFC beat frequency mitigated; \tabularnewline
 & & &  fiber agitator repaired \tabularnewline
5 & 2019-08-04 & - & realignment of FEM; \tabularnewline
& & &  replacement of LFC PCF
\tabularnewline
\bottomrule
\end{tabular}
\end{table}

The inputs to EXPRES are a $33 \times 132 \um$ science fiber as well as a $60 \times 180 \um$ extended fiber. The extended fiber is wider than the science fiber in the cross-dispersion direction, allowing for higher SNR flat-fielding for all pixels illuminated by light from the science fiber, particularly at the cross-dispersion edges. Both fibers are also supplemented by their own square simultaneous fibers, though these are not used in normal operation.

The flat field light source is provided by a custom solution: a collection of 25 LEDs integrated on a single compact chip. The wavelength range of the LEDs cover the entire bandwidth of EXPRES, and the relative power of each LED is tuned to approximately match the inverse response of the spectrograph. Light from these LEDs, averaging 12.5 W, is coupled into an integrating sphere and subsequently injected into both the extended and science fibers.

Wavelength calibration is carried out with a Menlo Systems laser frequency comb \citep[LFC; similar to those in][]{steinmetz_laser_2008,probst_relative_2016}. Three LFC exposures are taken through the science fiber every 15-30 minutes throughout a night.  Two ThAr exposures are taken each night, once at the beginning and once at the end. The LFC is set to standby between each set of calibrations to reduce wear on nonlinear optical elements, and is only turned on approximately one minute before the next exposure is needed, to suppress turn-on transients.

Barycentric corrections are derived from the EXPRES exposure meter, a $R\approx100$ spectrograph with an EMCCD detector and a bandpass that covers the spectral range of the LFC. During each science observation, the exposure meter takes a continuous series of 1 s exposures. Further technical details can be found in \citet{blackman_measured_2019}.

\hypertarget{analysis-of-expres-data}{%
\section{Analysis of EXPRES Data}\label{analysis-of-expres-data}}

\begin{figure*}
\centering
\input{figures/pipeline.tex}
\caption{Data flow for the EXPRES pipeline. Small black boxes represent different kinds of exposures (if in blue boxes) or different data associated with an exposure (if in green boxes). Double-boxes indicate finished data products, ready for use in subsequent science analysis.\label{fig:flowchart}}
\end{figure*}
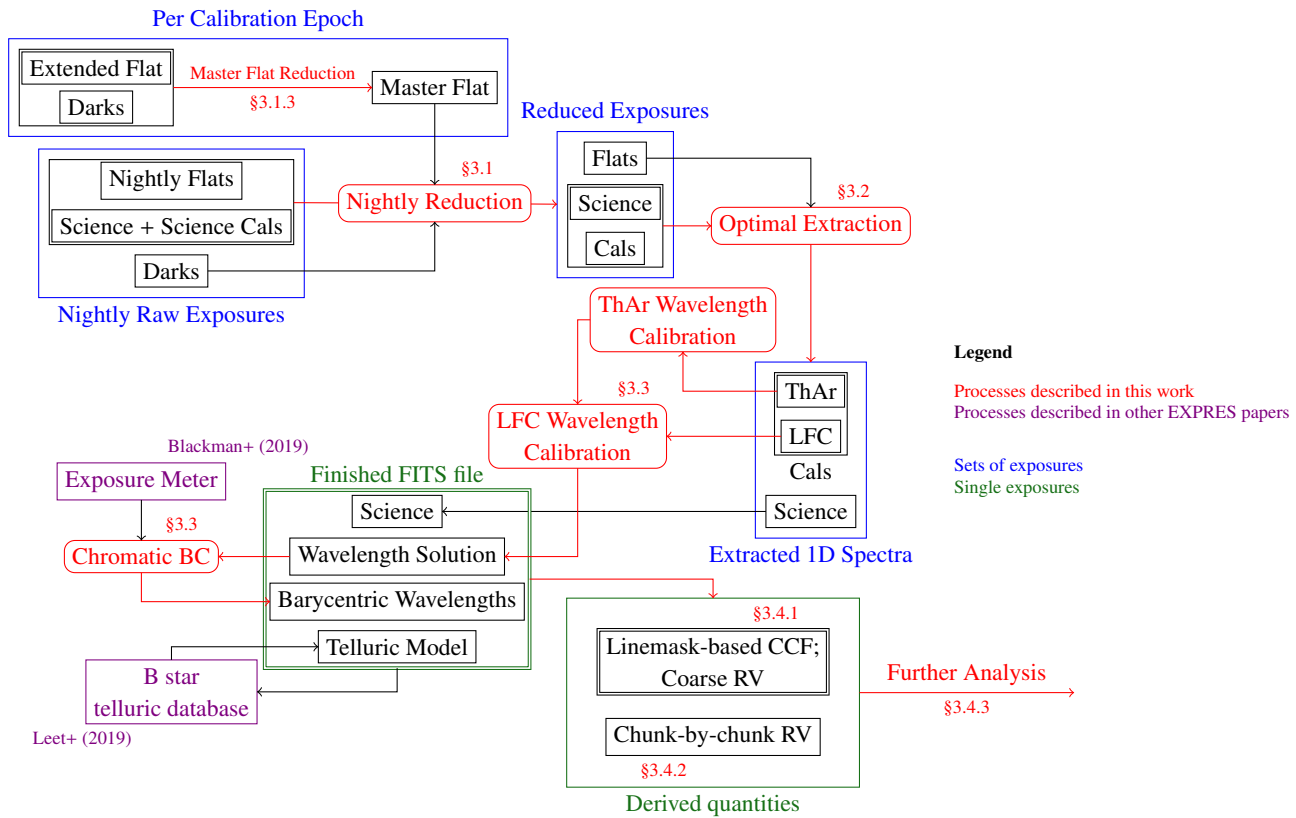

The EXPRES pipeline is written in Python and makes heavy use of the SciPy stack \citep{scipy}. We show a schematic representation of it in \cref{fig:flowchart}. In brief, the following steps are taken:

\begin{itemize}
\tightlist
\item At the start of each calibration epoch, several hundred extended flat images are taken. These are used to construct a master extended flat-field image, which is divided out from all exposures in the corresponding calibration epoch.
\item Each night, 30 dark and 30 science flat images are taken. They are used to reduce and extract the science frames taken the same night.
\item Echellogram orders are traced using the reduced science flats, a scattered light model is removed, and a flat-relative optimal extraction is performed. 
\item Wavelength solutions are interpolated for all science frames using bracketed LFC exposures, seeded by a nightly Thorium Argon source, as a calibration reference.
\item Telluric lines are identified empirically with SELENITE \citep{leet_tellurics_2019} and labelled for later analysis.
\item For exposures marked for RV analysis, we obtain radial velocities with two methods: cross-correlation against a line mask and a forward model.
\end{itemize}

We describe each of these steps in some detail below.

\hypertarget{reduction}{%
\subsection{Reduction}\label{reduction}}

The EXPRES detector is an STA1600LN CCD backside-illuminated image sensor with a $10,560 \times 10,560$ array containing $9 \um \times 9 \um$ pixels. The CCD is divided into 16 equal $5280 \times 1320$ pixel sections (two rows of eight) each with their own independent output amplifier and, thus, corresponding gain (see Table \ref{tab:gain}). Further information about the EXPRES CCD can be found in \citet{blackman_performance_2020}.

In the context of the EXPRES pipeline, ``reduction'' refers to the conversion of these 16 independent regions read in analog-to-digital units (ADU) to a full two-dimensional frame in units of photoelectrons. The reduction steps are as follows, dependent on the type of image being reduced (science, dark, science flat, or extended flat):
\begin{enumerate}
    \item Subtract a bias frame constructed from the overscan regions (all).
    \item Multiply each amplifier region by the corresponding gain coefficient (all).
    \item Median combine calibration frames (dark, science flat, extended flat).
    \item Subtract the reduced dark image (science, science flat, extended flat).
    \item Divide by the reduced master extended flat (science, science flat).
    \item Approximate the noise model using photon (Poisson) and read noise (science, science flat).
    \item Trace the echelle orders (science flat).
    \item Approximate the scattered light using a two-dimensional b-spline model (science, science flat).
\end{enumerate}

In the following subsections, we go into detail about each of the above steps. Information about the reduction of exposure meter data can be found in \citet{blackman_measured_2019}.

\subsubsection{Overscan}\label{overscan}

\begin{figure}
    \centering
    \includegraphics{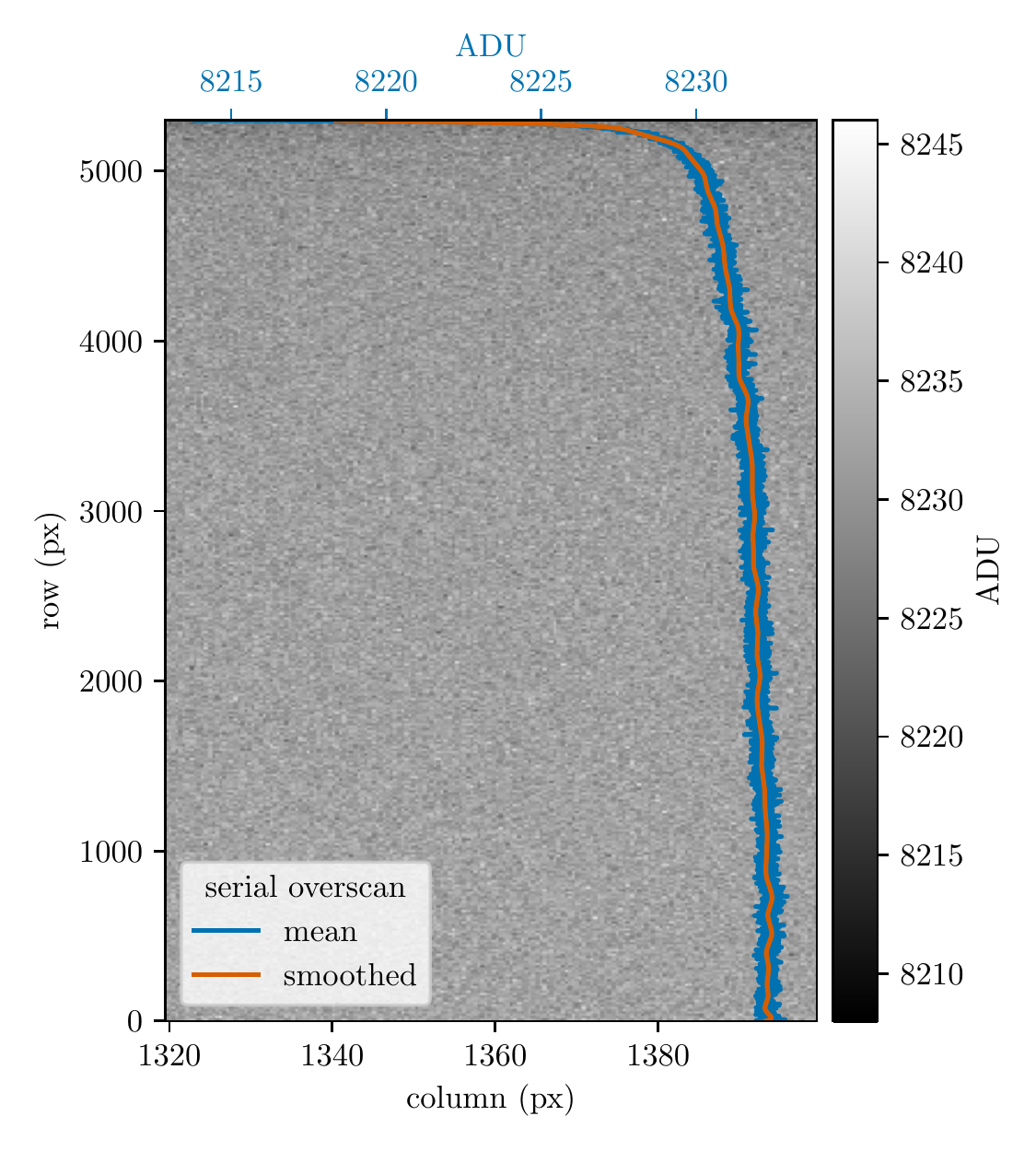}
    \includegraphics{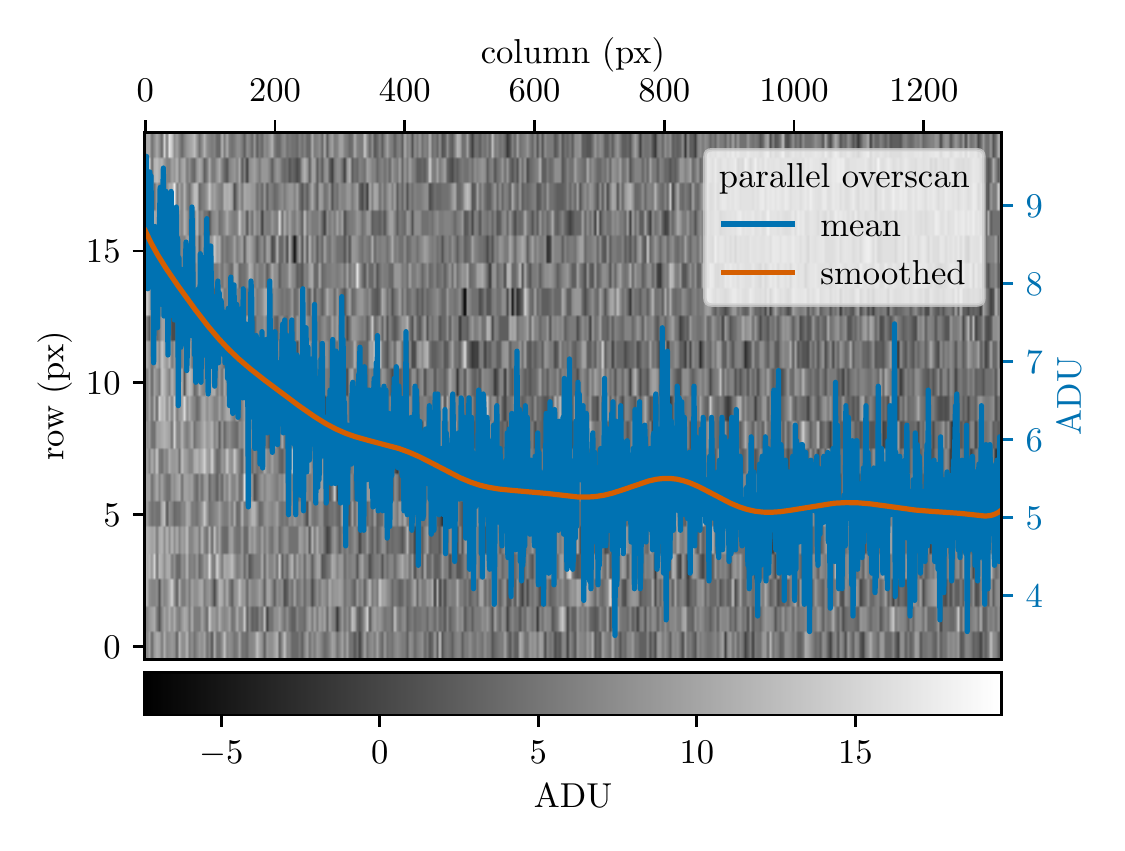}
    \caption{Bias approximation using the serial and parallel overscans. The mean and subsequent smoothed mean for each overscan are shown in blue and orange, respectively. Note that the counts shown in the parallel overscan were first subtracted by the first 20 rows of the serial overscan smoothed mean, generating values close to zero.}
    \label{fig:overscan}
\end{figure}

Each of the EXPRES CCD amplifier regions have overscans along both the serial (horizontal) and parallel (vertical) registers. The serial overscan is $5300 \times 180$ pixels along the right side of each amplifier region and the parallel overscan is $20 \times 1320$ pixels along the center line of the CCD (at the bottom of the upper regions and the top of the lower regions). Because these overscans contain virtual pixels read out by the same electronics as the real amplifier region, the EXPRES reduction pipeline uses them to approximate the bias of the CCD.  The process is as follows:
\begin{enumerate}
    \item Calculate the mean of the serial overscan region along its horizontal axis.
    \item Smooth this mean using a cubic b-spline with knots every $\sim 100$ pixels.
    \item Correct the rows in the parallel overscan region by subtracting the overlapping smoothed serial overscan region.
    \item Calculate the mean of the parallel overscan region along its vertical axis.
    \item Smooth this mean using a cubic b-spline with knots every $\sim 100$ pixels.
    \item Construct a bias for each pixel in the amplifier region by summing the corresponding row from the serial overscan and column from the parallel overscan.
\end{enumerate}
The mean and subsequent spline fit for the overscan regions of a single amplifier region are shown in \cref{fig:overscan}. This process is executed for all exposures, including darks.

\subsubsection{Gain}\label{gain}

The independent amplifier gains for the EXPRES CCD (Table \ref{tab:gain}) were determined empirically by matching the edges of neighboring amplifier regions based on stacked bias-subtracted extended flat images. We median combine 20 columns along each edge of two adjacent amplifier regions and determine the factor that minimizes the difference between them. We repeat this process along each row of amplifier regions, yielding a set of relative corrections for each row. Then, using the 20 median-combined rows along the center line of the detector, we determine the single factor that relates the top row of corrected amplifier regions to the bottom row.

Since these corrections are merely relative, we assume that the mean gain of the 16 amplifier regions given by the manufacturer is approximately correct. Thus, we match the mean of the empirically determined gain corrections to that given by the manufacturer. We note that the gains given by the manufacturer were not sufficient at matching the boundary conditions of the EXPRES CCD: some gain corrections were tweaked by more than 3\%.

\begin{table}[ht!]
\scriptsize
\centering
\caption{EXPRES CCD Gain\label{tab:gain}}
    \begin{tabular}{ccc|ccc}
        \toprule
        Amplifier & Empirical & STA & Amplifier & Empirical & STA \tabularnewline
        \midrule
        0 & 2.57980 & 2.6645 & 8 & 2.69787 & 2.6945 \tabularnewline
        1 & 2.55171 & 2.5352 & 9 & 2.65100 & 2.6422 \tabularnewline
        2 & 2.53844 & 2.5218 & 10 & 2.64354 & 2.6367 \tabularnewline
        3 & 2.52444 & 2.4065 & 11 & 2.60344 & 2.5502 \tabularnewline
        4 & 2.51480 & 2.6024 & 12 & 2.60497 & 2.5571 \tabularnewline
        5 & 2.49382 & 2.5686 & 13 & 2.59183 & 2.5630 \tabularnewline
        6 & 2.52745 & 2.4816 & 14 & 2.62881 & 2.5691 \tabularnewline
        7 & 2.53478 & 2.4960 & 15 & 2.72938 & 2.6111 \tabularnewline
        \bottomrule
    \end{tabular}
\end{table}

We were not able to calculate the gains in the typical manner---relating the variance of each pixel to its mean---because our flat-fielding LED source has an intrinsic variability that cannot be modeled by photon noise alone. See \citet{blackman_performance_2020} for further details.

\subsubsection{Master Extended Flat}\label{master-flat}

In order to measure the quantum efficiency (QE) variations with high signal across the relevant areas of the CCD, EXPRES employs an ``extended flat'' fiber. This rectangular fiber has slightly larger dimensions of $60 \times 180 \um$---as compared to the ``science'' rectangular fiber with dimensions $33 \times 132\um$---and is aligned with the center of the science fiber using a motor-driven mirror. Periodically throughout the operation of EXPRES, and at least once per epoch (Table \ref{tab:epochs}), the flat-fielding LED---typically used for order tracing and optimal extraction slit modeling---is injected into the extended flat fiber and a series of more than 100 images are taken of the resultant spectrum.

Using a median combination of these extended flat images, we construct a master extended flat-field image by dividing out a smooth fit to its echellogram. For each column of each order, we fit a parametric slit function: the squared convolution of a rectangle function with a Gaussian function, which has the analytic form
\begin{equation}
    P_{x,y}(A,d,\sigma) = A \left[ \Phi \left( \frac{y+d/2}{\sqrt{2}\sigma} \right) - \Phi \left(\frac{y-d/2}{\sqrt{2}\sigma} \right) \right]^2
    \label{eq:psf}
\end{equation}
where $\Phi$ is the error function, $d$ is the width of the rectangle, $\sigma$ is the standard deviation of the Gaussian, and $A$ is the amplitude.

This choice of functional PSF form was motivated by the physical nature of EXPRES and Fourier optics approximations. The input near field of the spectrograph is a rectangle function generated by the rectangular fiber. Taking the Fourier transform of the near field yields an approximation for the far field, which is subsequently morphed as it travels along the optical path of the instrument. We approximate this morph as a Gaussian function. Therefore, the resultant near field that is captured by the detector is the inverse Fourier transform of the modified far field. Invoking the convolution theorem, this detected near field is simply the convolution of the input rectangular function with a Gaussian function, which we subsequently square to approximate the total energy of the light as it hits the detector, yielding Equation \ref{eq:psf}.

Finally, we smooth the best-fit values for $A$, $d$, and $\sigma$ along each order using a cubic spline with ten equally spaced knots. The parameters from this smooth fit yield the profile with which we can divide the original extended flat image to generate the master extended flat-field template.

\begin{figure}
\centering
\includegraphics{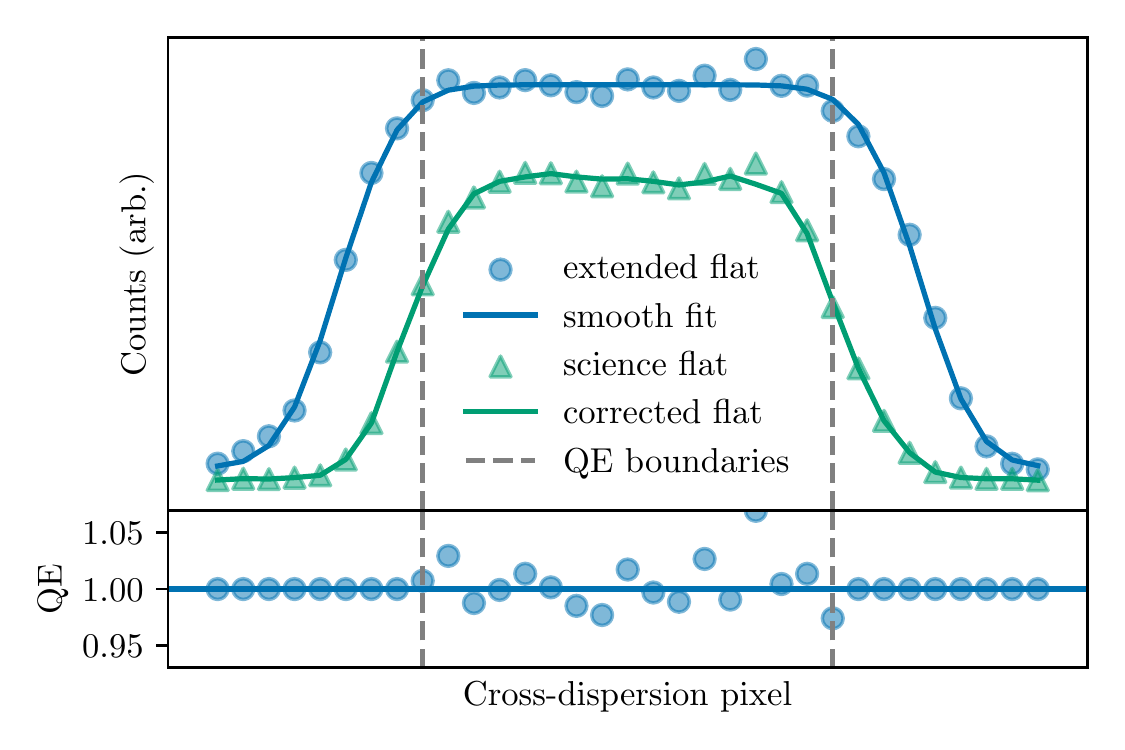}
\caption{Example cross-dispersion, cross-section of an extended flat as compared to a science flat. The best fit of \cref{eq:psf} is given for the extended flat. The resultant quantum efficiency (QE) corrections are shown in the lower plot and the bounds of the corrected region are demarcated. Note that these are corrections \textit{relative} to the mean QE. The QE corrected science flat is also shown in the upper plot.}\label{fig:flats}
\end{figure}

We only use pixels along the approximately flat portion of the extended slit in this constructed smooth profile. This decision was made for two reasons:
\begin{itemize}
    \item there is less signal along the top and bottom edges of the slit, thus there is inherently more scatter in the flat corrections for these pixels, and
    \item even with the analytic function that we use, the steep edges of the slit function are difficult to fit, which leads to systematic problems in the master extended flat.
\end{itemize}
Since the slit function of the extended flat was designed to be only 50\% larger than the science fiber slit function, the range of pixels covered by the flat portion of the extended slit does not correct the entire cross-dispersion profile of the science order. Thus, we also change the motor position of the extended flat injection mirror, which moves the extended flat slit function along the cross-dispersion direction of the echellogram. We thereby expand the number of pixels included in the master extended flat by generating a master extended flat image for each mirror position and then mean-combining these images. This process is completed for each epoch.

\subsubsection{Noise Model}\label{noise-model}

The two largest contributions to the noise model for any given pixel on the EXPRES CCD are photon noise and read noise, where these two quantities are measured and summed in quadrature for each pixel. Photon noise is assumed to be Poisson, such that the standard deviation is equal to the square root of the photoelectron counts. Read noise, on the other hand, is calculated empirically for each amplifier. First, the standard deviation of the nightly stack of dark frames is determined for every bias-subtracted gain-corrected pixel. Then, the median of these standard deviations is assigned as the read noise for each amplifier region. We assume that the read noise is consistent throughout each night of observation.

For the median-combined science flat frame, there is an additional noise term: intrinsic variability in the flat-fielding LED. As described in \citet{blackman_performance_2020}, the LED brightness varies by about 0.5\% over time. The uncertainty from this variability is proportional to the counts (as opposed to photon noise, which is proportional to the square root of the counts).  Therefore, it too is summed in quadrature to produce the total noise for the science flats. As with all median- or mean-combined frames, the noise model for the science flat is divided by the square root of the number of frames.

\subsection{Spectral Extraction}\label{spectral-extraction}

``Extraction,'' in the context of the EXPRES pipeline, refers to the process of converting reduced two-dimensional CCD data into a series of one-dimensional, normalized spectra, one for each order of the echelleogram. This process involves tracing the echelle orders, removing scattered light, executing the optimal extraction, and continuum normalizing the resultant spectra. Details for these steps are found in the following sections.

\subsubsection{Order Tracing}\label{order-tracing}

The orders of the echellogram are traced using the reduced science flat frame. First, the orders are detected using a peak-finding algorithm along the mean-combined center three columns. Then, for each order, moving from this center-line outward one column at a time, triplets of neighboring columns are mean-combined and the centroid of the resultant array is calculated. The right and left ends of each order are determined by setting a SNR > 30 threshold and stopping the trace once 50 subsequent columns do not reach this threshold. Finally, these centroids are smoothed along each echelle order using a 6th degree polynomial. A single set of traces is calculated for each night of observations.

\begin{figure}
    \centering
    \includegraphics{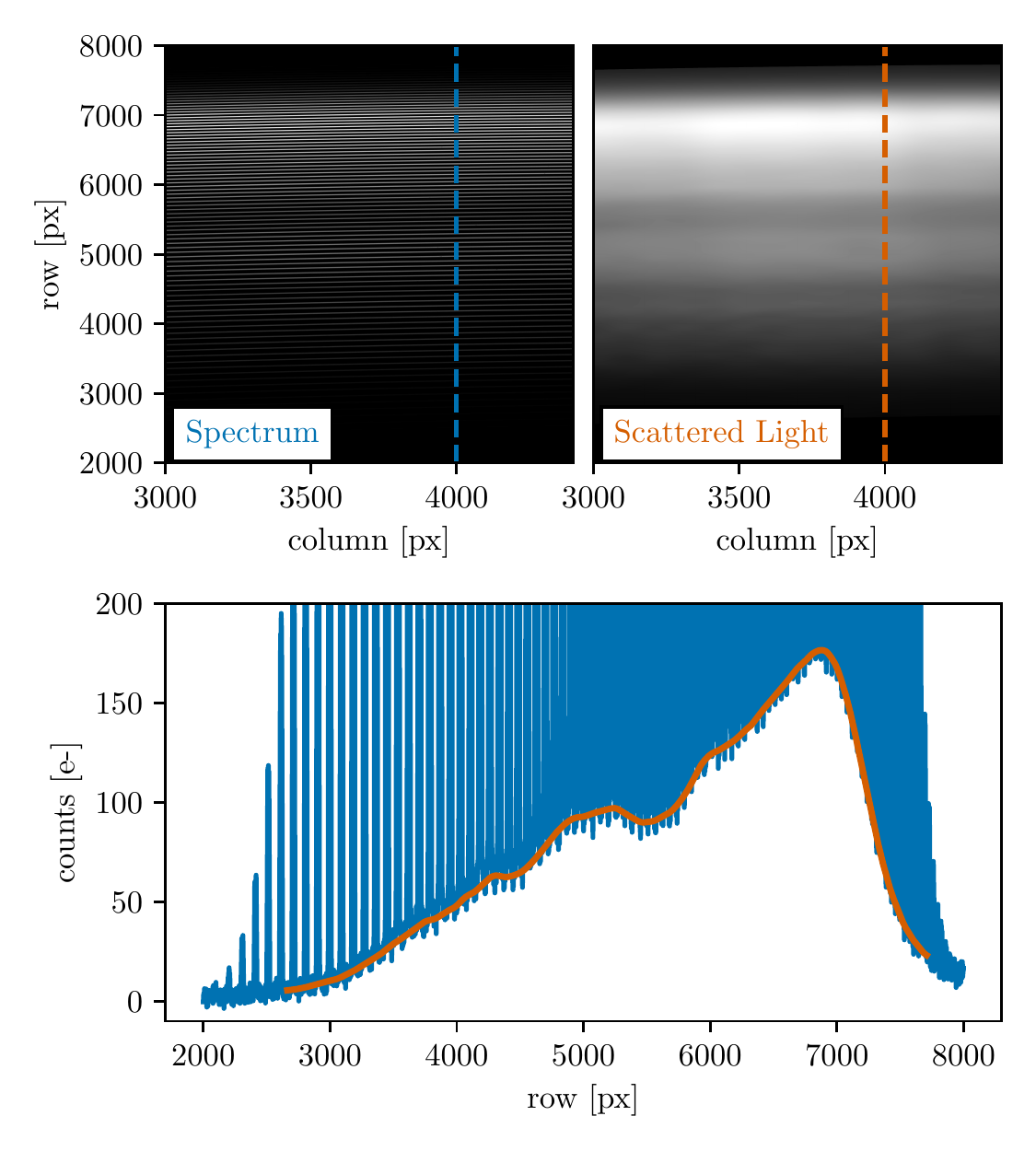}
    \caption{In the upper plot, two-dimensional scattered light approximation for reduced and median-combined science flat frames taken on October 24, 2019 are shown. The scales for the two images are not matched in order to empahsize the scattered light. The cross section of a single column (4000) for both the spectrum and calculated scattered light is shown in the lower plot. Note that the range of traced orders sets the upper and lower limits of the calculated scattered light.}
    \label{fig:scattered_light}
\end{figure}

\subsubsection{Scattered Light}\label{scattered-light}

Due to imperfections in the EXPRES optics, some scattered light from the instrument hits the CCD. This diffuse, scattered light is assumed to be smoothly varying across the detector and is estimated from the counts in the regions between the orders of the echellogram. First, a variance-weighted mean and associated uncertainty is calculated for each column of each inter-order region (including those immediately above and below the traced echellogram) using the seven pixels set halfway between adjacent traced orders. These inter-order background approximations are then smoothed using a cubic b-spline with knots set every $\sim$100 columns. Cosmic rays that could potentially skew this smoothed fit are iteratively rejected using a $5\sigma$ outlier cut.

The two-dimensional scattered light image is generated through a quadratic interpolation along each column of the smoothed inter-order backgrounds. The calculated scattered light is subtracted from the associated reduced image before any further extraction. This process is completed for the science flats, stellar frames, and all wavelength calibration frames. An example of a subregion of the resultant scattered light approximation is shown in \cref{fig:scattered_light}.

\subsubsection{Optimal Extraction}\label{optimal-extraction}

\begin{figure*}
    \centering
    \includegraphics{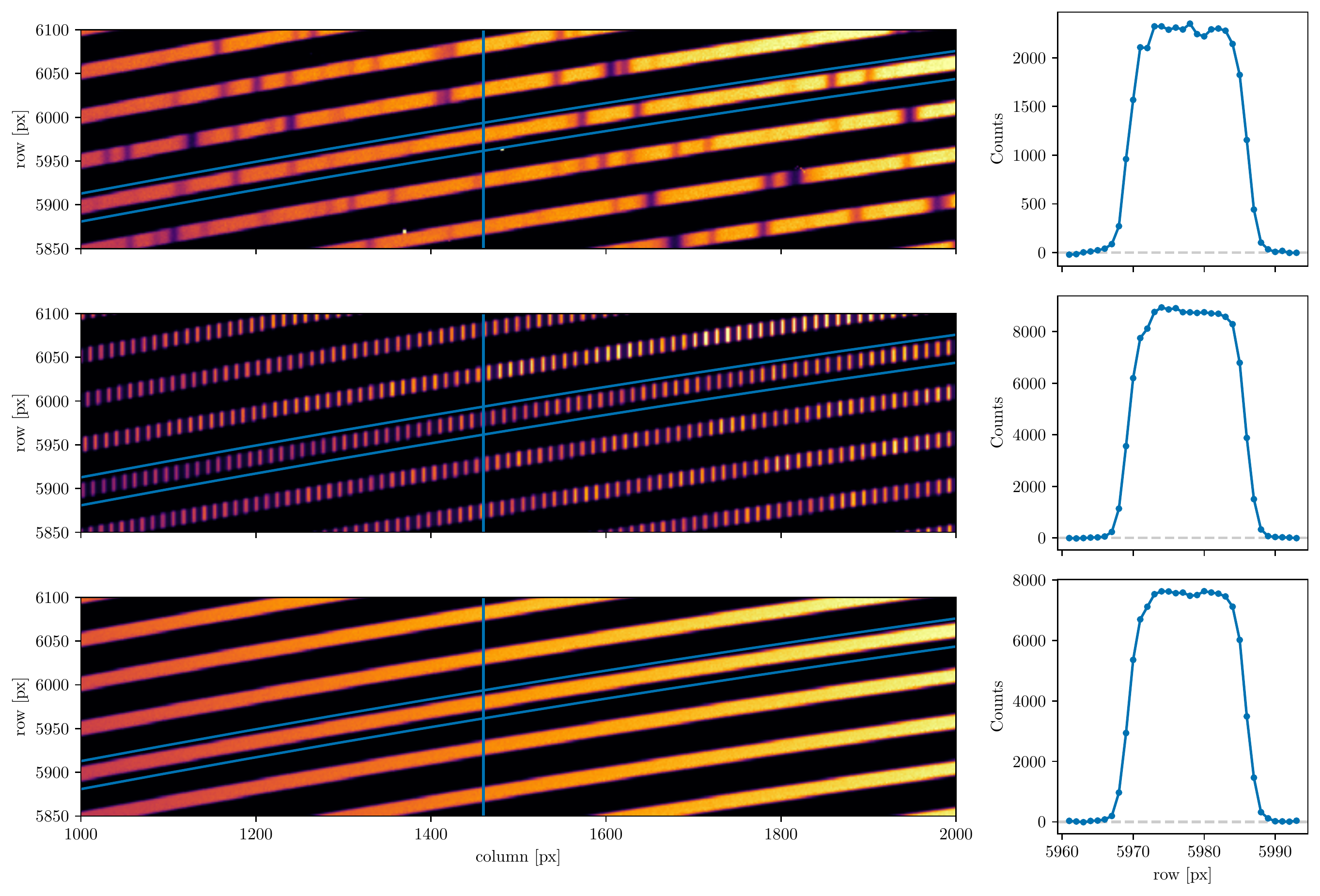}
    \caption{Section of raw images for a science exposure (above; HD 217014), laser frequency comb (middle), and calibration flat (below) taken with EXPRES on 24 October 2019. The extraction aperture for our optimal extraction of echelle order 100 is shown on the flat images as the intersection of the traced order (with vertical extent of 33 pixels) and the single-column slit at $x = 1460$, both shown in blue. The reduced counts in the extraction aperture for each image (i.e. removing hot pixels and QE variations) are shown in the right panels, as a function of pixel row position $y$.}
    \label{fig:extraction}
\end{figure*}

The optimal extraction algorithm that we implement in the EXPRES pipeline is an updated version of the algorithms developed by \citet{horne_optimal_1986,piskunov_algorithms_2002,zechmeister_flat_2014}. Following along the traces (from Section \ref{order-tracing}) of each echelle order, we construct a least-squares estimator for each 33-pixel tall column $x$:
\begin{equation}
    \chi_x^2 = \sum_y ( D_{x,y} - P_{x,y} s_x )^2  w_{x,y}
    \label{eq:least-squares}
\end{equation}
where $D_{x,y}$ are the photoelectron counts for each pixel $(x,y)$ in the reduced data, $P_{x,y}$ is the model of the slit function corresponding to those same pixels, $s_x$ is the extracted spectral intensity, and $w_{x,y}$ are the weights for each pixel. These weights are inversely proportional to the variance of $D_{x,y}$ ($\sigma_{x,y}^2$, see Section \ref{noise-model}) and include a binary cosmic ray mask, $M_{x,y}$, i.e. $w_{x,y} = M_{x,y} / \sigma^2_{x,y}$ (see \citealt{zechmeister_flat_2014}). The minimization of \cref{eq:least-squares} has an analytic solution:
\begin{equation}
    s_x = \frac{\sum_y w_{x,y} D_{x,y} P_{x,y} }{\sum_y w_{x,y} P_{x,y} P_{x,y} }.
    \label{eq:least-squares-soln}
\end{equation}
The propagated uncertainty of the extraction
\begin{equation}
    \sigma_{s_x} = \sqrt{\frac{1}{\sum_y w_{x,y} P_{x,y} P_{x,y}}}
\end{equation}
is rescaled by $\chi_{\mathrm{red,}x}$ as in \citet{zechmeister_flat_2014}. However, we smooth $\chi_{\mathrm{red,}x}$ across each order using a 3rd-order polynomial before applying it to $\sigma_{s_x}$ to avoid low number statistic variance.

Cosmic rays are rejected by iteratively adding to the cosmic ray mask, $M_{x,y}$, based on a tiered outlier rejection algorithm. After calculating $s_x$ for an entire order, the residual \((D_{x,y}-P_{x,y}s_x)w_{x,y}\) for each pixel is calculated. For each column in the order with a pixel that exceeds $8\sigma$, the pixel with the largest residual is rejected and $s_x$ is re-calculated for that column. This process repeats until all $8\sigma$ outliers are rejected. Next, this same rejection process repeats for $2\sigma$ outliers that also neighbor previously rejected pixels, thus rejecting the dim tails of otherwise bright cosmic rays. Importantly, this second step is executed using two-dimensional pixel information, meaning that a bright cosmic ray on a pixel in a given column can have its tail rejected in an adjacent column.

The EXPRES pipeline has two distinct methods of approximating the model $P_{x,y}$ in \cref{eq:least-squares-soln}:
\begin{enumerate}
    \item fit each column of the science flat with a parametric slit function (\cref{eq:psf}), smooth the parameters along each order using a b-spline, and then normalize the function; or
    \item use the science flat, without normalization, as in \citet{zechmeister_flat_2014}.
\end{enumerate}
This yields two modes of operation when extracting data, wherein (1) keeps the echelle blaze function of each order intact while (2) intrinsically removes the blaze.

We note that using method (2) requires inclusion of the variance prior in the flat \(\sigma_{P_{x,y}}^2\), as well as that for the data \(\sigma_{D_{x,y}}^2\), when determining the weights \(w_{x,y}\) of the least-squares estimator. Therefore, the variance prior for each pixel should instead be constructed as
\begin{equation}
    \sigma^2_{x,y} = \sigma^2_{D_{x,y}} + s^2_x \sigma^2_{P_{x,y}},
\end{equation}
and an additional nonlinear cost term must be added to \cref{eq:least-squares} to prevent $s_x \rightarrow \infty$. In our case we choose \(\sum_y \ln{\sigma^2_{x,y}}\).

Thus, the minimization of \cref{eq:least-squares} must be solved numerically:
\begin{equation}
    s_x^{(n+1)} = \frac{\sum_y{ w^{(n)}_{x,y} D_{x,y} P_{x,y}  }}{\sum_y{ w^{(n)}_{x,y} P_{x,y} P_{x,y}} + R^{(n)}_x}
    \label{eq:numerical-soln}
\end{equation}
where the weights $w^{(n)}_{x,y}$ for each iteration are calculated using the previous iteration's solution $s^{(n)}_x$ and $s^{(0)}_x$ is calculated assuming all $\sigma^2_{P_{x,y}} = 0$. The relaxation factor $R^{(n)}_x$ is also chosen to be
\begin{equation}
    R^{(n)}_x = \sum_y \sigma^2_{P_{x,y}} w^{(n)}_{x,y} - \sum_y \sigma^2_{P_{x,y}} \left( w^{(n)}_{x,y} \right)^2 \left( D_{x,y} - s_x^{(n)} P_{x,y} \right)^2
    \label{eq:relaxation}
\end{equation}
from the minimization of the modified \cref{eq:least-squares}. This numerical process is repeated until a relative tolerance of $10^{-10}$ is met. Thus, method (2) is operationally slower than method (1), but intrinsically accounting for un-corrected QE variations and automatically removing the blaze without relying on a blaze model outweigh this minor increase in computational cost.

\begin{figure}
    \centering
    \includegraphics{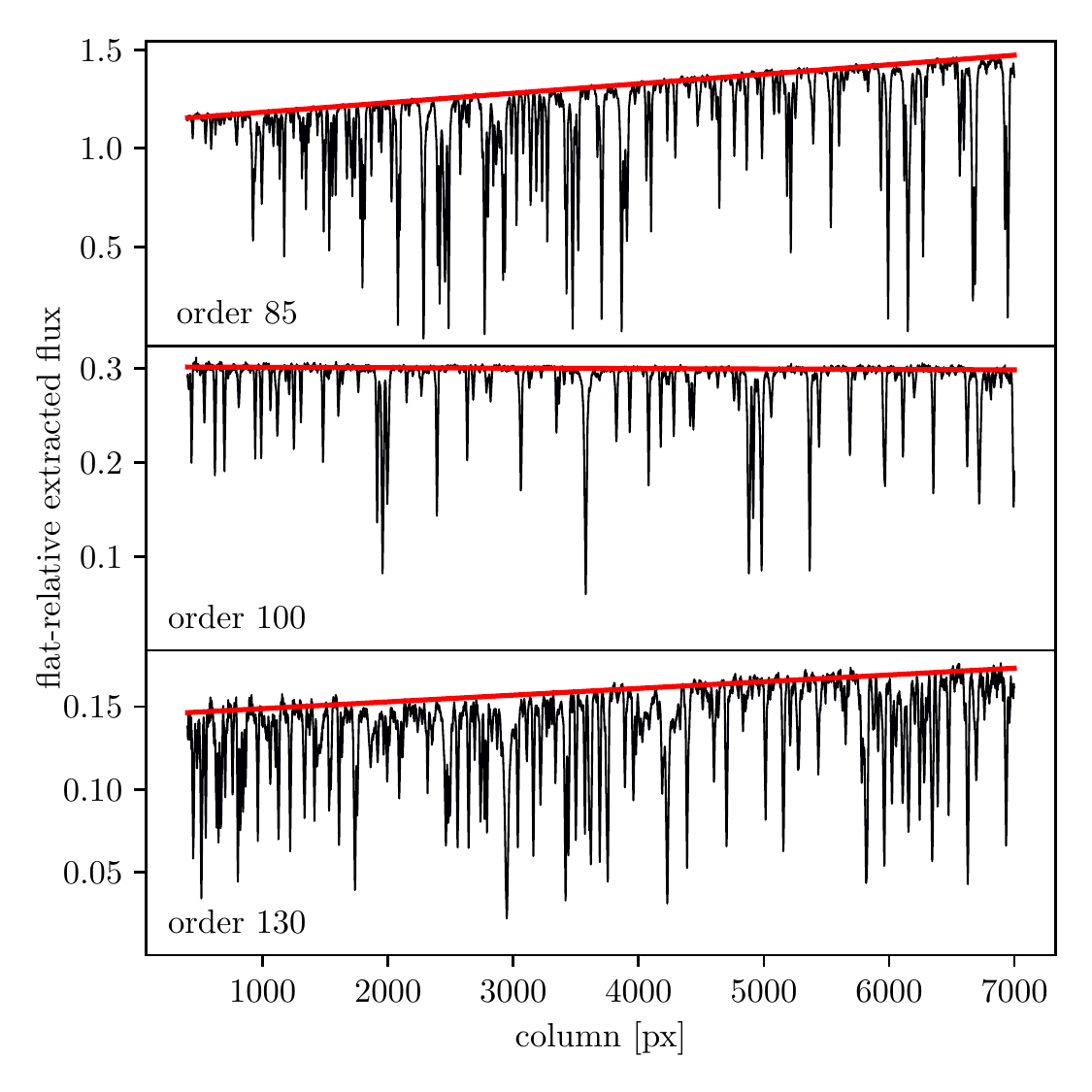}
    \caption{Flat-relative optimally extracted flux (black) and associated calculated continua (red) for three orders from HD217014 observed on 24 October 2019. All continua were calculated using a linear model. Notice that the spectral flux relative to the flat-fielding LED source flux is significantly different between orders, but the resultant continuum across each order varies slowly enough to remain essentially linear.}
    \label{fig:continuum}
\end{figure}

Removal of the blaze through method (2) can be naively understood by recognizing \cref{eq:least-squares} is approximately solved by $s_x \approx \frac{D_{x}}{P_{x}}$, where $P_x$ is the spectral intensity of the LED source times the blaze and $D_x$ is the spectral intensity of the star times the blaze. Both $P_x$ and $D_x$ describe their respective intensity at the given column $x$ of the order. Therefore, the resultant continuum of a stellar spectrum extracted using method (2) is simply the continuum of the star divided by the spectrum of the LED source. Since both of these are slowly varying functions, we can model the extracted stellar continuum with a simple linear model for each order, calculated iteratively with $2\sigma$ outlier rejection for those pixels contained in absorption lines. See \cref{fig:continuum} for examples of extracted spectra and the approximated continua.

In the EXPRES pipeline, we strictly use (2) to extract all data and only use (1) as a secondary check for our extracted RVs. However, we note that maximizing the SNR of the cross-correlation function described in Section \ref{radial-velocity-solutions} requires weights, which increase with signal-to-noise of the stellar continuum, to be assigned to different portions of the spectrum. Unfortunately, the posterior uncertainties of the extraction have too much scatter to provide these weights. We find that using the blaze function is a good proxy for a smooth weighting function; therefore, we approximate a blaze function model for each epoch using the $\alpha$-hull method \citep{xu_alpha_hull_2019} applied to a science flat extracted with (1). We find that RV results obtained with (2) including the separately defined blaze function are superior to those results from spectra extracted with (1).

Extraction method (2) is also the most appropriate choice for wavelength calibration spectra, such as those from the ThAr lamp and the LFC. The calibration methods used in Section \ref{wavelength-calibration} are linear and can make use of the scattered posterior uncertainties as weights when fitting each emission line without systematically shifting results as in cross-correlation. Additionally, light from the LED, ThAr lamp, and LFC all travel through approximately the same lengths of fiber, meaning removal of an instrumental response function (i.e. a calibration ``continuum'') should not be necessary when using method (2).

The signal-to-noise ratio (SNR) of a given observation is reported here as the maximum $s_x / \sigma_{s_x}$ in echelle order 111. This is effectively the per-pixel SNR at 550 nm, conforming to the metric used by \citet{fischer_eprv_2014} to compare many contemporaneous spectrographs. The resolution element of EXPRES contains approximately 4 pixels \citep{jurgenson_expres_2016}, thus the per-resolution-element SNR is simply twice the per-pixel SNR.

\hypertarget{wavelength-calibration}{%
\subsection{Wavelength Calibration}\label{wavelength-calibration}}

EXPRES uses a laser frequency comb (LFC) as its primary calibration source, which generates a series of spectral lines evenly spaced in frequency, whose nominal frequencies $\nu_n$ satisfy the relation
\begin{equation}\nu_n = \nu_\text{rep} \times n + \nu_\text{offset},\label{eq:lfc}\end{equation}
for integers $n$. The repetition rate $\nu_\text{rep}$ and offset frequency $\nu_\text{offset}$ are referenced against a GPS-disciplined quartz oscillator, providing calibration stability corresponding to a fractional uncertainty of less than $8\times10^{-12}$ for integration times greater than 1\unit{s}.

While it is possible to obtain an absolute calibration from the LFC once the free spectral range of the echellogram has been adequately characterized, the LFC suffers from poor throughput in very blue and very red orders. In particular, although our instrumental throughput is sufficient to permit order tracing and extraction from echelle orders 75 through 160 (for $3800\unit{\AA}<\lambda<8220\unit{\AA}$), the LFC only sufficiently illuminates echelle orders 82 through 135 (for $4500\unit{\AA}<\lambda<7500\unit{\AA}$). The photonic crystal fiber (PCF) of the LFC was then replaced in July 2019 due to the decreasing stability of the LFC. With the replacement, the polarization of the LFC was switched, making the LFC redder and thus changing the orders illuminated by the LFC to echelle orders 82 through 130 (with the blue edge at \(5300\unit{\AA}\)). The polarization switch should significantly increase the lifetime and stability of the LFC. Consequently, we use ThAr lamp exposures taken at the beginning and end of each night as a secondary calibration source to provide well-constrained wavelength solutions for orders outside the range of the LFC.

Calibration triplets (3 LFC's) are taken through the science fiber at roughly 15-30 minute intervals throughout observing, interwoven with science exposures. The exposure times of these calibration frames are chosen to match the target SNR of the science exposures. While EXPRES is equipped with a secondary square fiber to permit simultaneous wavelength calibrations, we choose to take calibrations through the science fiber so that our calibration data sample the same pixels and optical elements as the science exposures.  This strategy aims to homogenize our exposures to pixel-level, uncalibratable systematic errors. Also, as shown by \citet{blackman_performance_2020}, the instrumental stability of EXPRES is such that sampling the LFC every 15-30 minutes provides enough information to correct for any instrumental changes throughout the night, as simultaneous calibration would. Calibration images are taken while the telescope is slewing and so typically cost little additional time (less than 2 minutes an hour).

A ThAr wavelength solution is generated from each ThAr exposure using the IDL code \texttt{thid.pro}, developed by Jeff Valenti, which identifies ThAr lines by matching lines in an exposure against a line atlas. A 6th-order, 2D polynomial is then fitted over pixel location \(x\) and the absolute echelle order \(m\) against the scaled wavelength \(m\lambda\). Matching lines against an atlas is performed manually once at the beginning of each calibration epoch; otherwise, the wavelength solution from the immediately preceding ThAr exposure is used as an initial guess for the locations of atlas lines in a given ThAr exposure, allowing this process to be automated. Since the LFC lines are sparse relative to the precision of the ThAr calibration (1 LFC line every 10 pixels on average with the ThAr calibration accurate to the nearest pixel), this is sufficient to permit unambiguous mode identification for the LFC lines.

For any given LFC exposure, the locations of modes are identified by fitting Gaussians to each peak after a smooth background has been subtracted. An initial, trial wavelength solution is generated by linearly interpolating the ThAr solutions from the beginning and end of the night. These are used to determine the mode number \(n\) corresponding to the frequency of each mode. Once again, a 2D polynomial is fitted for \(m\lambda\) as a function of \(m\) and \(x\). Since the LFC produces a far denser set of lines---typically about 20,000 lines across 50 orders are identified in an LFC exposure, compared to about 4,000 lines across 82 orders in a ThAr exposure---we use a 2D polynomial described by a 10x10 matrix of coefficients (9th-order in each dimension) for the fit. The locations of the ThAr lines are also included in the fit in order to constrain the behaviour of this polynomial in echelle orders that are otherwise inaccessible to the LFC.

\begin{figure}[htbp]
    \centering
    \includegraphics{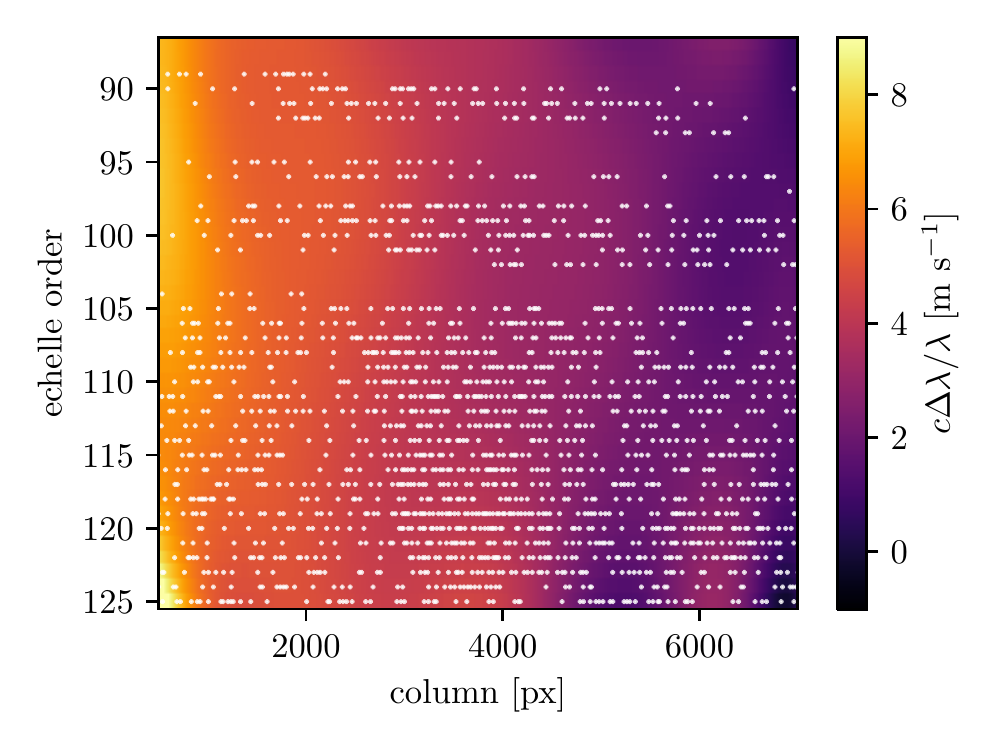}
    \caption{Calibration drift over the course of a single night (24 October 2019) for the LFC-constrained region of the echellogram, plotted in terms of the absolute echelle order number and the pixel column of the CCD. The average calibration drift for the whole night ($\sim$4.0\ms) is of similar magnitude to local variations in the drift, therefore using a single average velocity offset would necessarily incur significant additional calibration error. In other words, the wavelength solution along the left and right edges of the shown spectral format would be offset by -4.0 and +4.0\ms respectively if a single velocity offset were to be used. Spectral lines used for radial velocity solutions (shown as white dots drawn from the ESPRESSO G2 linelist) sample the detector in a nonuniform fashion, and result in different overall velocity offsets depending on spectral type.}
    \label{fig:caldrift}
\end{figure}

For each of the polynomial coefficients describing the wavelength solution, we fit a smooth function (a cubic polynomial) in time. We have found that interpolating the polynomial coefficients, rather than directly interpolating the pixel-wise wavelength solutions, is more robust to imperfections in individual calibration frames. This set of 100 functions is evaluated at the photon-weighted midpoint time of each science exposure to generate a wavelength solution. This differs from the standard practice at other spectrographs (e.g. HARPS and ESPRESSO), where a single velocity offset, rather than a time-dependent wavelength solution, are assigned to each science exposure. We choose to do this in order to accommodate time-dependent variations in the characteristics of the instrument, which may lead to calibration shifts that cannot be adequately described by a single average velocity offset.

We illustrate this in \cref{fig:caldrift}, where we show the calibration drift over the course of a single night in the LFC-constrained region of the echellogram. Our radial velocity solutions (see Section \ref{radial-velocity-solutions}) sample spectral lines that are non-uniformly distributed across the detector; therefore, it is advantageous to characterize local variations to prevent overall systematic offsets. Of course, the precise differential velocity imparted by this calibration drift ultimately depends on the spectral type as well as both barycentric and systemic velocity of the star under consideration.

As the final step in wavelength calibration, the EXPRES pipeline applies a barycentric correction to the wavelength solution of each stellar observation using the method described by \citet{blackman_accounting_2017}. For each 1~s exposure of the exposure meter, a barycentric correction is calculated using BARYCORR \citep{wright_barycorr_2014}. The photon-weighted average barycentric correction is calculated for each of 8 wavelength bins of the spectra. A third-degree polynomial is then fit to these averages, yielding a smoothly-varying wavelength-dependent barycentric correction \(z_\text{B}(\lambda)\) for the observation. As shown by \citet{tronsgaard_bc_2019}, it is important to distinguish this from a photon-weighted midpoint time used to calculate an overall chromatically-dependent barycentric correction \citep[e.g.,][]{landoni_bc_2014} as this can impart a $\sim$10\cms systematic error to the radial velocity especially in cases of longer exposure times, high airmass, or poor seeing.

Finally, we apply \(z_\text{B}\) directly to the wavelength solution:
\begin{equation}
    \lambda_\text{bary} = \lambda_\text{lab}^\text{(vac)} \left(1 + z_\text{B}\left(\lambda_\text{lab}^\text{(air)}\right)\right)
    \label{eq:bary_wavelength}
\end{equation}
where \(\lambda_\text{lab}\) is the LFC-generated lab-frame wavelength solution and \(\lambda_\text{bary}\) is the wavelength solution in the frame of the solar system barycenter. Because the EXPRES exposure meter is not in vacuum (as opposed to EXPRES itself), \(z_\text{B}(\lambda)\) is measured using air-wavelengths. Therefore, \(\lambda_\text{lab}\) is converted from vacuum to air using the algorithm and parameters derived by \citet{ciddor_vactoair_1996} before applying the barycentric correction in \cref{eq:bary_wavelength}.

\hypertarget{radial-velocity-solutions}{%
\subsection{Radial Velocity Solutions}\label{radial-velocity-solutions}}

The data analysis pipeline of EXPRES employs two distinct computational techniques to independently extract radial velocities from stellar spectra:
\begin{enumerate}
    \item a ``cross-correlation function'' method \citep[CCF; see][]{baranne_ccf_1979} is used to determine a rough estimate of the absolute radial velocity for each observation (Section \ref{cross-correlation}), and
    \item a forward model based on a morphed NSO solar spectrum is used to derive a more precise relative radial velocity curve (Section \ref{forward-modeling}).
\end{enumerate}
Both of these methods are currently implemented in the EXPRES pipeline for self-validation. Simultaneous results from both methods are presented in Section \ref{initial-results}. The methods as implemented in the EXPRES pipeline are described as follows.

\subsubsection{Cross-Correlation}\label{cross-correlation}

As the first step in our analysis, a CCF method estimates the absolute RV of EXPRES science targets precise to several tens of \cms (depending on photon noise). We also use the CCF method to diagnose drifts and instabilities in our calibration sources, using line lists given by the comb parameters following \cref{eq:lfc} for the LFC or a ThAr line atlas for the ThAr lamp.

The CCF is constructed from the input spectrum \(f(\lambda)\) as well as a spectral-type linelist---a set of spectral lines at rest vacuum wavelengths \(\left\{\lambda_i(0)\right\}\) associated with contrast weights \(\left\{c_i\right\}\) and widths \(\left\{h_i\right\}\). For a given trial radial velocity \(v\), the wavelength of each line in the linelist is redshifted appropriately to
\begin{equation}
    \lambda_i(v) = \lambda_i(0) \sqrt{c + v \over c - v}.
\end{equation}
The CCF is then computed as a numerical approximation to the integral
\begin{equation}
    \mathrm{CCF}(v) = \int \mathrm{d} \lambda f(\lambda) \sum_i c_i w\left(\lambda - \lambda_i(v) \over h_i\right)
    \label{eq:ccf}
\end{equation}
where \(w\) is an aribtrary window function approximating a Dirac \(\delta\) function and \(\lambda\) is with respect to the barycentric-corrected wavelength solution from \cref{eq:bary_wavelength}.

The CCF in \cref{eq:ccf} is computed independently for each echelle order with a variety of trial velocities, and the CCFs for all relevant orders are co-added before a velocity model is fitted. This is a similar practice to other CCF-based RV pipelines \citep[e.g.~][]{brahm_ceres_2017}. When deriving extreme precision radial velocities, we only include orders falling within the spectral range of the LFC, since in principle it affords considerably better sampling density and calibration stability than those regions covered by the ThAr lamp alone.

An appropriate functional model is then fitted against the co-added values of the CCF. The position parameter and posterior uncertainties of the fitted model are returned as the reported velocity and formal errors. Other quantities of astrophysical interest (e.g.,~rotational broadening width, bisector inverse slope) are also computed from the co-added CCF.

Our construction of the CCF incorporates the ability to use an arbitrary window function \(w\). In the current iteration of the EXPRES pipeline, we use a cosine function, matching other contemporary CCF implementations \citep[e.g.~][]{espresso_drs_2013, espresso_drs_2019, brahm_ceres_2017}. We also use a Gaussian functional model to fit the CCF for our reported radial velocities. Other possible combinations of window functions and CCF models are discussed in Section \ref{further-analysis}.

\subsubsection{Forward modeling}\label{forward-modeling}

In addition to our CCF RV solution, we have developed a new forward-modeling technique by adapting algorithms developed for the iodine RV technique \citep{marcy_iodine_1992, butler_iodine_1996} as well as ideas from the ``line-by-line'' method developed by \citet{dumusque_measuring_2018}. Forward-modelling from empirical stellar spectral templates is known to produce velocities with less statistical scatter than the CCF method, and typically measures relative rather than absolute radial velocities \citep{anglada_harps_2012}. Our modeling process is simplified relative to the iodine method because the optical design of EXPRES was optimized to provide stability in the line spread function (LSF) of the spectrograph \citep{jurgenson_expres_2016, blackman_performance_2020}, eliminating the need to model the instrumental LSF with several free parameters. In addition, free parameters for wavelength solution and dispersion are eliminated since the barycentric wavelength solution (\cref{eq:bary_wavelength}) is provided as part of the nightly optimal extraction.

First, we construct a spectral template for each stellar target. An ideal template will have very high SNR and will be a good spectral match to the program stars. Our starting point is to obtain a set of four consecutive spectra---each with SNR of about 250---providing an effective SNR of 500 per pixel or SNR of 1000 per resolution element. As described by \citet{dumusque_measuring_2018}, we prefer to use spectra with low barycentric velocities so that the program spectra shift around the approximate zero point wavelengths. The telluric contamination is then modeled in each spectrum using SELENITE \citep{leet_tellurics_2019} and divided out. Finally, the set of spectra are co-added.

\begin{figure}
    \centering
    \includegraphics[trim=5cm 3cm 3cm 4cm, clip]{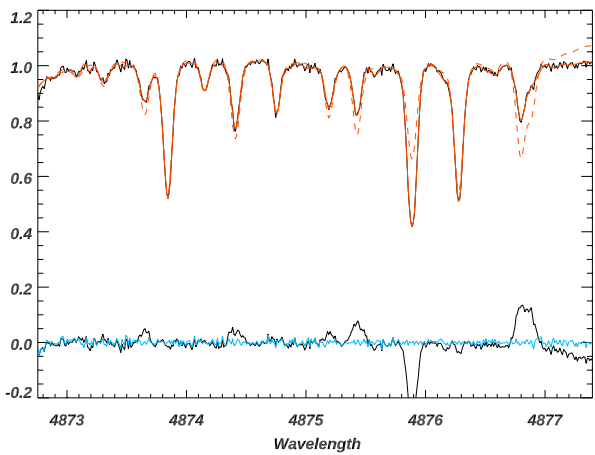}
    \caption{The high-resolution, high-SNR NSO spectrum (red dashed line) is shifted and cubic-spline interpolated to the wavelength scale of a program observation (upper black line). The difference spectrum (bottom black line) is used to identify discrepancies between the spectra above the photon noise threshold (bottom blue line). A Levenburg-Marquardt algorithm drives the growth of pseudo-lines until the NSO spectrum has morphed to the match the spectrum of the program star (solid red line). This morphed spectrum is then used as a template for forward modeling.}
    \label{fig:morph}
\end{figure}

However, even the co-added spectrum will not provide a high enough SNR for a robust template. Therefore, we take the additional step of morphing the NSO solar spectrum (see \cref{fig:morph}) with a native SNR $\sim$10,000 and resolution $\sim$500,000 to match the co-added, telluric-cleaned spectra for each of our program stars using the following procedure:
\begin{enumerate} 
\item The co-added program star spectrum is divided into $\sim$2000 individual chunks that are 140 pixels ($\sim $2\AA) wide within the orders of the spectrum covered by the LFC.
\item The barycentric wavelengths of the program star are used to extract a segment of the NSO spectrum with generous padding of 200 pixels for modeling shifts. This segment of the NSO spectrum is shifted to the barycentric frame of the co-added program star spectrum.
\item A Levenburg-Marquardt (L-M) least squares algorithm is used to (i) determine the best-fit width for a Gaussian convolution kernel to rotationally broaden the NSO spectrum, (ii) refine the Doppler shift of the NSO spectrum, and (iii) apply a vertical shift to align the continuum of the NSO and the co-added spectrum. 
\item The rotationally broadened and shifted NSO spectrum is cubic-spline interpolated onto the wavelength scale of the co-added spectrum. 
\item A difference spectrum is calculated. Nodes are dropped down consecutively at points where the absolute value of the difference spectrum exceeds a threshold, characterized by the photon noise of the co-added spectrum. Pixels with the largest residuals in the difference spectrum are modeled first. The maximum number of nodes is 60, but depending on the chunk there are typically about a dozen nodes required to model the NSO spectrum for each 130-pixel chunk. 
\item At each node a positive or negative Gaussian feature with a width characterized by the line spread function of EXPRES is used to perturb the NSO spectrum; the depth of the morphing feature is determined by L-M fitting of the residuals. 
\item Iterative growth of the morphing lines stops when the residuals of the difference spectrum are consistent with photon noise.
\item Each chunk of the template is weighted according to the amount of spectral information, using the SNR and the derivative of intensity $I$ with pixel:
\begin{equation}
    \sum{\frac{\delta I \ \lambda}{c \ \delta \lambda} \frac{1}{\mathrm{SNR}}}
\end{equation}
\end{enumerate}

Once the template for each star has been generated, a L-M fit is executed for every 140-pixel chunk of the program star spectra. Telluric-affected pixels in each observation are assigned zero weight in the fit. There are only two free parameters for each chunk: a Doppler shift and continuum normalization scale factor. Thus, each chunk---45 LFC-calibrated orders each with about 47 chunks yielding $\sim$2,000 total chunks---provides an independent measurement of the relative RV for the star. The RV for each chunk is subsequently subtracted by the mean of that chunk over all observations, thus removing any offsets that might occur because of geometric anomalies in the detector while preserving the spread in RV variations.

Weights for each chunk are determined using empirical arguments, the $\chi^2$ of the L-M fit, and a chunk-specific modifier based on its relative temporal scatter. Chunks that do not contain any absorption lines in the stellar template as well as chunks that yield relative velocities greater than $\pm$1000\ms are assigned zero weight. Moreover, chunks that have $\chi^2 > 5.0$ (typically occurring if an incorrect stellar template was used or a telluric line was missed, for example) and remaining chunks that are among those with the largest 3\% of reduced $\chi^2$ are all assigned zero weight.

Because some chunks have less spectral information, there will be more scatter in the RVs derived from these chunks. For example, chunks in the blue part of the spectrum typically have several absorption lines, but chunks in the red part of the spectrum may have only one spectral line meaning the L-M fitting will not be well-constrained. Likewise, telluric contamination within a given chunk can manifest as large scatter in the RV over time. Therefore, the non-zero weight for given chunk $i$ within an observation $j$ is assigned as
\begin{equation}
    w_{i,j}^{-1} = \chi^2_{i,j}~\frac{\sum_{j}^{n} (v_{i,j} - \bar{v}_{j})^2}{n-1}
    \label{eq:chunk-weight}
\end{equation}
where $\bar{v}_{j}$ is the median velocity for all chunks of observation $j$ and $n$ is the total number of observations for a given stellar target. The reported RV measurement for each observation is thus a weighted mean of the individual chunk velocities and the formal error is the corresponding standard error of the weighted mean.

\hypertarget{further-analysis}{%
\subsubsection{Further analysis}\label{further-analysis}}

Once RVs have been derived, the extracted spectra and CCFs are passed down the pipeline for more sophisticated analysis. The spectral range of EXPRES is intended to permit characterisation of stellar activity and planetary atmospheric absorption lines. For chromospheric activity in particular, we extract the Ca\,\textsc{ii} line core emission ratio index $S_\mathrm{HK}$ \citep[using the parametric model of][]{isaacson_chromospheric_2010}, calibrated to yield results consistent with the Mt. Wilson Observatory catalogue \citep{duncan_stellar_1991}.

We also aim to incorporate spectroscopic activity indicators directly into the RV solution methodology. For example, in addition to using a rectangular ``box'' function and truncated cosine in the CCF, which are implemented in other similar velocity analysis codes \citep[e.g.~][]{espresso_drs_2013, espresso_drs_2019, brahm_ceres_2017}, the EXPRES analysis code implements CCF computation using Gauss-Hermite window functions of the form
\begin{equation}w(x) = {1 \over \sqrt{2^n n! \sqrt{\pi}}}H_n(x) \exp\left[-{x^2 \over 2}\right],
\end{equation}
where \(H_n\) is the \(n\)\textsuperscript{th} (physicists') Hermite polynomial. Computing higher-order CCFs as coefficients in a Hermite-functional decomposition, and more generally with respect to different orthogonal basis functions, will permit more sophisticated analysis of stellar activity (via a sparse description of variations in the CCF line profile), as an alternative parameterization to current derived observables (such as the CCF bisector inverse slope/FWHM). Alternatively, data-driven decorrelation of stellar activity from bulk radial velocities, or alternative template-based RV solution methodologies (Holzer et al., in prep), may be possible once we have built up an archive of stellar spectra.

\section{Initial Results - HD 217014}\label{initial-results}

We now present velocities derived with this pipeline based on 47 observations of HD 217014 (51 Peg) over the ten month span between the beginning of Epoch 4 and December 1, 2019. We do so to examine various sources of uncertainty and error in the velocimetric pipeline, while avoiding the known instrumental instabilities inherent in Epochs 1-3 \citep[see][for details]{blackman_performance_2020, szymkowiak_lfc_2019}, and to compare the two radial velocity methods outlined in Section \ref{radial-velocity-solutions}. Observations with an SNR less than 160 are not included in this analysis. We construct our CCF using the G2 linemask from the ESPRESSO pipeline \citep{espresso_drs_2013, espresso_drs_2019} and a cosine window function. We also fit all RVs with a single planet Keplerian model, constrained by the literature value of the orbital period \citep[4.2308 days,][]{wang_eccentricity_2011}. This Keplerian model is parameterized by the velocity semi-amplitude ($K$), the eccentricity ($e$), the argument of periastron, and a phase of periastron.

\begin{figure}[htbp]
    \centering
    \includegraphics{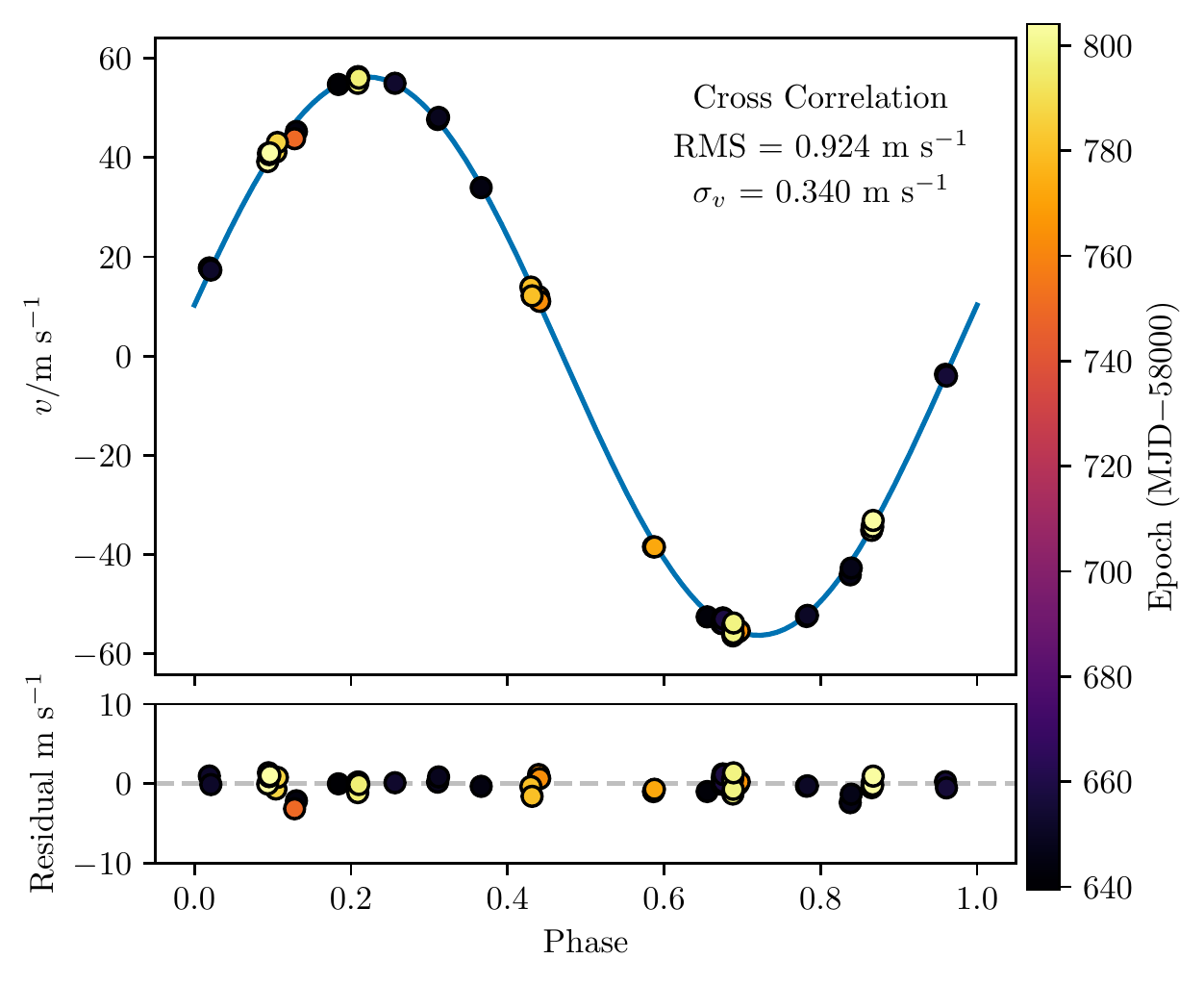}
    \includegraphics{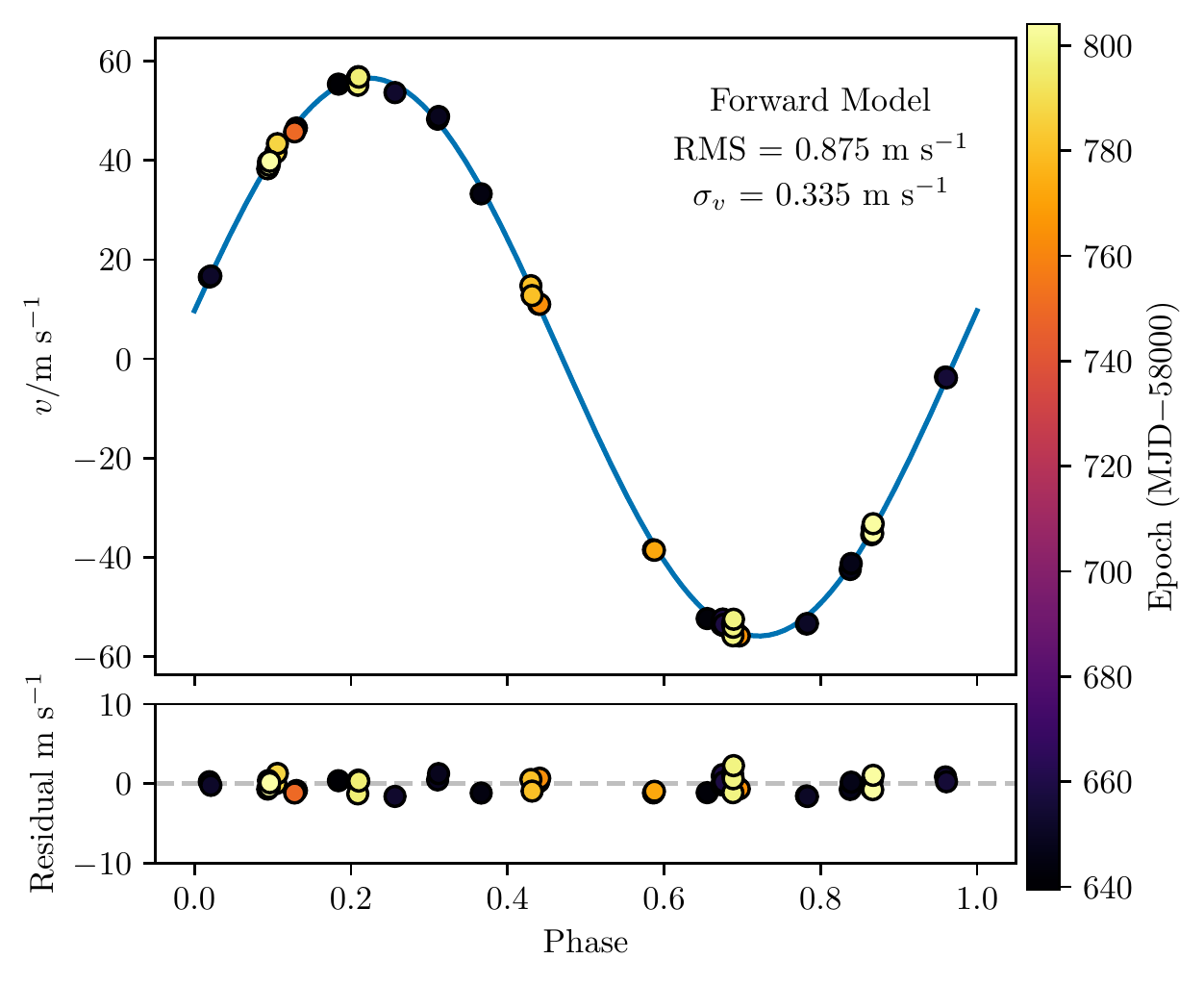}
    \caption{Phased radial velocities, Keplerian orbital fits, and residuals for EXPRES observations of 51 Peg b. The figure is labelled with the RMS residual to the fitted orbital solution, as well as the median formal error $\sigma_v$ of all data points shown.}
    \label{fig:217014}
\end{figure}

\begin{table}[ht!]
    \footnotesize
    \centering
    \caption{EXPRES commissioning RVs of 51 Peg (full data set available online)
    \label{tab:51pegvels}}
    \begin{tabular}{ccccc}
        \toprule
        BMJD & $V_{ccf}/\cms$ & $V_{fm}/\cms$ & RV Epoch & SNR
        \tabularnewline
        \midrule
        58639.45844 & $-3320697\pm20$ & $ 5739\pm32$ & 4 & 385 \tabularnewline
        58641.45174 & $-3331424\pm44$ & $-5035\pm42$ & 4 & 179 \tabularnewline
        58643.46218 & $-3321644\pm34$ & $ 4854\pm34$ & 4 & 225 \tabularnewline
        58644.46095 & $-3322776\pm35$ & $ 3527\pm34$ & 4 & 233 \tabularnewline
        58646.45596 & $-3330577\pm39$ & $-4045\pm38$ & 4 & 203 \tabularnewline
         & \(\vdots\) & & \tabularnewline
        \bottomrule
    \end{tabular}
\end{table}

\cref{fig:217014} shows the resulting radial velocities from both the CCF and Forward Model methods along with their respective orbital fits. Because the CCF uses a linelist with absolute wavelengths, the fit systemic velocity of $-33.2603(5)\unit{km\,s^{-1}}$ \citep[in excellent agreement with][]{gaiadr2} has been removed. For each of the RV epochs, we also fit an independent velocity offset relative to the overall offset. For epochs 4 and 5, these are $-1.5(4)\ms$ and $1.2(5)\ms$ when using the CCF method, and $-1.2(8)\ms$ and $0.8(7)\ms$ when using the forward model. These offsets account for modifications to the instrumental systematics owing to the various fiber changes and realignments. The offsets differ slightly in magnitude between the two RV analysis methods because they intrinsically weigh regions of the detector differently, accentuating or mitigating certain instrumental systematics.

\begin{table}[ht!]
\footnotesize
\centering
\caption{Fit parameters for 51 Peg b\label{tab:51peg}}
\begin{tabular}{cccccc}
\toprule
Instr. & $K/\ms$ & $e$ & RMS$/\ms$ & $\sigma_v/\ms$ \tabularnewline
\midrule
EXPRES CCF & $56.24\pm0.14$ & $0.000\pm0.002$ & 0.924 & 0.340 \tabularnewline
EXPRES FM & $56.26\pm0.13$ & $0.007\pm0.003$ & 0.875 & 0.335 \tabularnewline
HARPS DRS & $53.4\pm1.6$ & $0.062\pm0.010$ & 0.941 & 1.023 \tabularnewline
HIRES & $56.7\pm0.4$ & $0.020\pm0.007$ & 2.74 & 1.169 \tabularnewline
\bottomrule
\end{tabular}
\end{table}

We show the values of the Keplerian fit parameters in \cref{tab:51peg} along with the RMS residual and the median formal error ($\sigma_v$). The parameter uncertainties were derived by taking the square root of the product of the posterior variances and reduced $\chi^2$ of the least squares fit. By way of comparison, we also perform the same procedure with eight years of archival velocities from the HIRES instrument on the Keck I telescope \citep[corrected for instrumental systematics per][]{talor_correcting_2019} and four months of data from the HARPS DRS \citep{trifonov_harps_2020}. The EXPRES orbital solution parameters for 51 Peg are consistent with those returned from these previous studies, but with higher precision due to the improved formal errors. The two EXPRES RV methods are also internally consistent, with the forward model producing a slightly more favorable RMS. Finally, we note that the RMS of the EXPRES fit residuals approximately matches that of the HARPS DRS, with the EXPRES fit returning parameters more comparable with literature values \citep{bedell_woble_2019,wilson2019,wang_eccentricity_2011}.

\hypertarget{discussion}{%
\section{Discussion}\label{discussion}}

Each step of the pipeline contains several free parameters — for instance, the degree of the fitting polynomials to use for the spatial wavelength solution fits and temporal smoothing, as well as various SNR threshholds for calibration line identification and order inclusion in the CCF.  To assist community users of the instrument, we have opted, as far as possible, to preselect reasonable default values for most of these parameters, which may be overridden at runtime. In what follows, we document some nonobvious but critical aspects of these systematics, and we illustrate some aspects of the decision-making process for choosing our default values for some of these parameters.

\hypertarget{formal-errors}{%
\subsection{Formal vs. true velocity errors}\label{formal-errors}}

Since the formal velocity errors returned from the CCF fitting procedure are constructed only from the co-added CCFs and their propagated errors, they do not account for effects like wavelength calibration error (inducing spurious velocity shifts) or time estimation error (via erroneous barycentric corrections). Instead, they mostly reflect velocity estimation error due to photon noise being propagated to the CCF.

On the other hand, the formal velocity errors returned from the forward model fitting procedure does include some information
about relative uncertainty in certain regions of the EXPRES detector. For instance, a chunk that tends to have a telluric line will naturally incur more spread in the measured RV for that chunk. Therefore, even though this chunk is down-weighted by our analysis, its spurious effect still propagates to the RV error.

\begin{figure}
    \centering
    \includegraphics{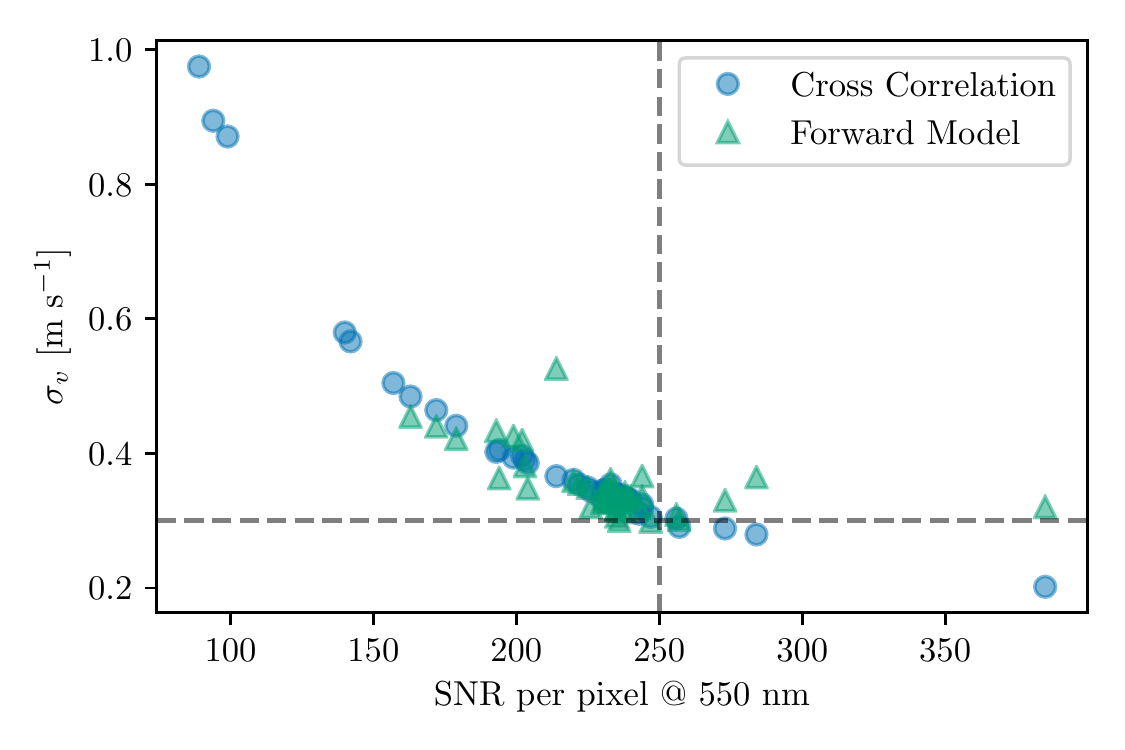}
    \caption{Formal RV errors returned from both RV fitting procedures, plotted against the per-pixel SNR (as a characterization of photon noise) for velocities from Section \ref{initial-results}. EXPRES's target SNR of 250 is shown with a vertical line and the approximate associated formal error of 30\cms is shown with a horizontal line. Note that data here with SNR less than 160 are not included in \cref{fig:217014}.}
    \label{fig:snr}
\end{figure}

These assumption are bourne out in \cref{fig:snr}, showing these formal errors as a function of the observation SNRs. The CCF points depend essentially only on photon noise and potentially CCD readout and optimal extraction systematics---which we are confident of having adequately accounted for---up to some constant that may depend on e.g.~the choice of CCF linemask or window function, or intrinsic astrophysical properties of the target. Conversely, the Forward Model formal errors contain much more scatter that we believe folds in some uncertainty from telluric contamination and, potentially, stellar noise. Thus, our estimation of the true photon noise limit of EXPRES is better described by the propagated errors of the CCF analysis, though the two analyses yield quite similar results. As shown in \cref{fig:snr}, we define this limit to be $30\cms$ for a single observation at the EXPRES target SNR of 250.

Following our diagnosis and repair of the LFC beat frequency noted in \cref{tab:epochs}, we measured the remaining uncalibratable velocity errors arising from wavelength calibration in particular to be relatively small---between 4 and 6\cms{} RMS (Szymkowiak et al., in prep). There also exist several other sources of error (e.g. from uncalibratable instrumental systematics and guiding errors), that constitute additional contributions to the RV error budget. To correctly estimate the velocimetric error, one needs to appropriately account for and then combine these error terms (e.g. by adding them in quadrature) with the formal value reported from the RV analysis. A detailed inventory of these error sources \citep{blackman_performance_2020} estimates the combined instrumental and guiding errors of EXPRES at $\sim$10\cms. Thus the single observation error of EXPRES is clearly dominated by the apparent photon noise.

\hypertarget{ccf-order-selection}{%
\subsection{Chromatic dependences}\label{ccf-order-selection}}

Many of the novel techniques that we have adopted in the EXPRES pipeline involve the introduction of chromatic dependences into quantities that have previously been considered to be uniform with wavelength, such as calibration offsets and the barycentric offset velocity. It therefore behooves us to investigate possible chromatic effects that emerge at the end of the CCF velocity-solving and orbit-fitting procedure.

The CCF analysis in Section \ref{initial-results} was performed by co-adding CCFs derived from echelle orders 126 through 86 ($4850\unit{\AA}<\lambda<7150\unit{\AA}$) before fitting an absorption-line model to derive a velocity. These orders are those for which at least $N_\text{min}=19$ LFC lines are detected that pass both the SNR threshhold and all quality checks imposed by our peakfitter ($N_\text{min}$ depends on the degree of the polynomial fitted to the wavelength solution, which is a free parameter in our code, as are the threshhold values for these quality checks).

\begin{figure}
\centering
\includegraphics{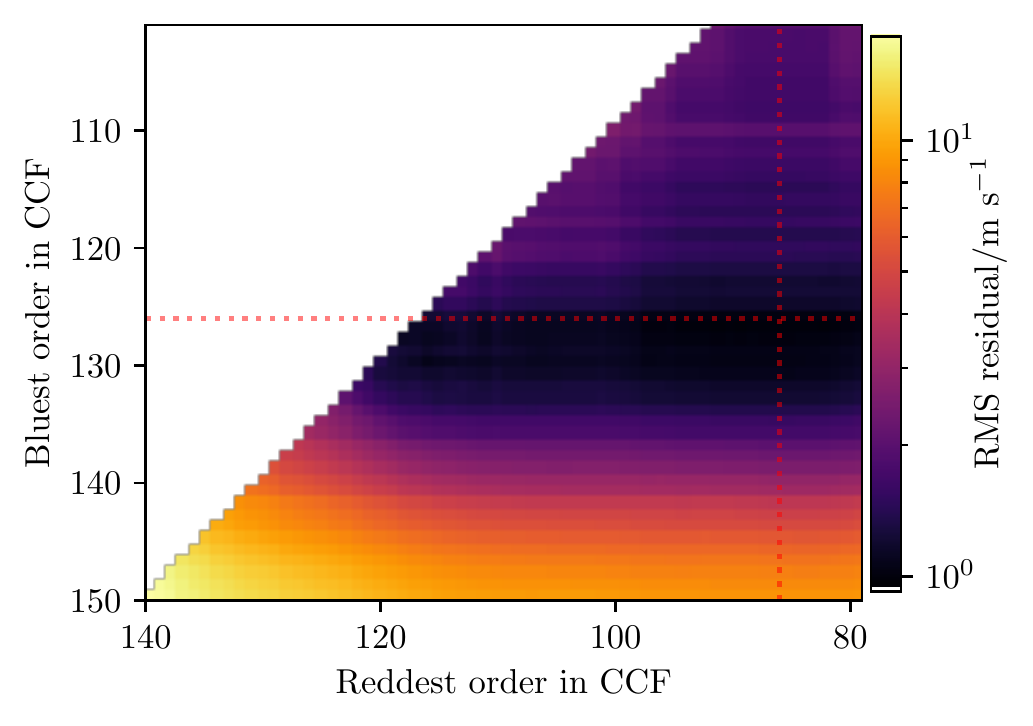}
\caption{RMS scatter from orbital solutions fitted to HD 217014 as a function of bluest and reddest echelle order included in the CCF computation. Points along each diagonal indicate sets of velocities computed with the same number of echelle orders. At least 10 echelle orders have been included in all CCF computations. Dotted red lines show the default parameters used in the preceding sections.\label{fig:orders}}
\end{figure}

In \cref{fig:orders}, we show the RMS residuals from the orbital solutions that arise when we repeat these analyses while varying the range of echelle orders used when co-adding CCFs; our default parameter selections are indicated with the red dotted lines. In particular, this means extrapolating the wavelength solution and chromatic barycentric correction beyond the spectral range of the LFC, which covers echelle orders 130 through 82, and of the chromatic exposure meter, which covers orders 135 through 86 ($\sim4650\unit{\AA}<\lambda<7150\unit{\AA}$).

For this data set, we see that there is a sharp dependence on the bluest order co-added into the CCF. Moreover, we see that the introduction of orders redder than the LFC cutoff also slightly increases the RMS error to the fit. Recalling that the wavelength solution outside of the LFC region is largely constrained by the ThAr lamp, this potentially implies a calibration offset between the LFC and the ThAr lamp, despite both sources illuminating the instrument through the same fiber.

On the other hand, we do not see any similarly sharp cutoff when extrapolating the chromatic barycentric correction to outside of the wavelength range covered by the exposure meter. This suggests that the wavelength dependences of our barycentric corrections (detailed more fully in \citealt{blackman_measured_2019}) are generally smooth enough for robust extrapolation.

Finally, it is possible to choose narrower ranges of echelle orders that yield smaller RMS errors than our default parameter selection. However, we note that this is potentially dependent on the specifics of the CCF linelists used in the computation and also possibly on underlying astrophysical properties of the science targets and other fortuitous factors. We have opted to include as many orders as can be accurate, so as to minimize photon noise, while still leaving the option to include fewer orders available to end-users.

\hypertarget{conclusion}{%
\section{Conclusion}\label{conclusion}}

The commissioning process on the EXPRES instrument is essentially complete, along with the development of an optimal extraction pipeline that we have been using for preliminary RV analysis through both CCF and forward modelling techniques. Within the instrumental back-end (i.e. limiting ourselves to the calibration unit and the spectrograph proper), we have determined our photon-noise-limited RV errors to be approximately 0.3\ms for a single observation with SNR of 250. With EXPRES's current observing strategy of four observations per night per target, this result implies a nightly measurement error of only 0.15\ms. While our on-star measurement error appears to be $\sim$0.9\ms---based on residual RMS to an orbital fit of 51 Peg b---we must also note that our RV analysis pipeline does not fully address photospheric velocity sources, telluric contamination, or longer-term instrumental errors. These other sources of RV scatter are beyond the scope of this paper although we are actively investigating them. The RV precision presented in this paper, therefore, represents our worst-case scenario in the absence of further improvements. We anticipate upcoming hardware improvements and more sophisticated RV solution methodologies to only enhance our measurement precision and long-term instrumental stability. 

Moreover, there are multiple parts of the pipeline that remain under active development. We are investigating the use of spectro-perfectionism \citep{bolton_specperf_2010, cornachione_specperf_2019} as an alternative to optimal extraction. We are also exploring a hierarchical, non-parametric wavelength solution that takes advantage of the low degrees of freedom allowed in a stabilized instrument and the density of lines offered by new wavelength calibrators (Zhao et al. in prep). There are also plans to implement a more data-driven approach (as in \citealt{bedell_woble_2019}, with modifications to permit chromatic barycentric corrections) as yet a third RV analysis technique. 

These caveats notwithstanding, we have demonstrated that the technical innovations that have been invested into the development of novel instrumentation and software analysis techniques for EXPRES have largely paid off — they have permitted us to unambiguously attain sub-\ms on-sky radial velocity precision. Presently, this makes EXPRES the most precise EPRV spectrograph in the northern hemisphere.

Finally, we hope the lessons we have learned in the process of commissioning this instrument, and the techniques we have developed, to be of some value to the community of EPRV instrument builders moving forward.

\acknowledgements{We thank the anonymous referee for comments that significantly improved this paper. We gratefully acknowledge Yale University, the Heising-Simons Foundation, and an anonymous Yale alumni donor for providing telescope time for EXPRES used to obtain the data in this paper. We also acknowledge the extensive work that went into the RePack spectral extraction code, written by Lars Buchhave; this code helped to benchmark the EXPRES optimal extraction described in this paper. JO thanks X. Dumusque for illuminating and productive discussions. We especially thank the NSF for funding that allowed for precise wavelength calibration and software pipelines through NSF ATI-1509436 and NSF AST-1616086 and for the construction of EXPRES through MRI-1429365. This material is based upon work supported by the National Science Foundation Graduate Research Fellowship under Grant No. DGE1122492 (RRP, LLZ, ABD).}

\software{SciPy library \citep{scipy}, NumPy \citep{numpy, numpy2}, Astropy \citep{astropy:2013,astropy:2018}.}

\facilities{LDT}

\bibliography{vels.bib}

\end{document}

%% file: meta.tex
\newcommand{\chinesename}{{\begin{CJK}{UTF8}{gbsn}(王加冕)\end{CJK}}}

\title{An Extreme Precision Radial Velocity Pipeline: First Radial Velocities from EXPRES}
\correspondingauthor{Debra Fischer}
\email{debra.fischer@yale.edu}


\author[0000-0003-2168-0191]{Ryan R. Petersburg}
\affiliation{Department of Physics, Yale University, 217 Prospect St, New Haven, CT 06511, USA}
\affiliation{Department of Astronomy, Yale University, 52 Hillhouse Ave., New Haven, CT 06511, USA}
\author[0000-0001-7664-648X]{J. M. Joel Ong \chinesename}
\affiliation{Department of Astronomy, Yale University, 52 Hillhouse Ave., New Haven, CT 06511, USA}
\author[0000-0002-3852-3590]{Lily L. Zhao}
\affiliation{Department of Astronomy, Yale University, 52 Hillhouse Ave., New Haven, CT 06511, USA}


\author[0000-0002-0303-3276]{Ryan T. Blackman}
\affiliation{Department of Astronomy, Yale University, 52 Hillhouse Ave., New Haven, CT 06511, USA}
\author[0000-0002-9873-1471]{John M. Brewer}
\affiliation{Dept. of Physics \& Astronomy, San Francisco State University, 1600 Holloway Ave., San Francisco, CA 94132, USA}
\affiliation{Department of Astronomy, Yale University, 52 Hillhouse Ave., New Haven, CT 06511, USA}
\author[0000-0003-1605-5666]{Lars A. Buchhave}
\affiliation{DTU Space, National Space Institute, Technical University of Denmark, Elektrovej 328, DK-2800 Kgs. Lyngby, Denmark}
\author[0000-0001-9749-6150]{Samuel H. C. Cabot}
\affiliation{Department of Astronomy, Yale University, 52 Hillhouse Ave., New Haven, CT 06511, USA}
\author[0000-0002-5070-8395]{Allen B. Davis}
\affiliation{Department of Astronomy, Yale University, 52 Hillhouse Ave., New Haven, CT 06511, USA}
\author[0000-0002-5018-7761]{Colby A. Jurgenson}
\affiliation{Department of Astronomy, The Ohio State University, 4055 McPherson Laboratory, 140 West 18th Avenue, Columbus, OH 43210, USA}
\affiliation{Department of Astronomy, Yale University, 52 Hillhouse Ave., New Haven, CT 06511, USA}
\author[0000-0001-5847-9147]{Christopher Leet}
\affiliation{Department of Astronomy, Yale University, 52 Hillhouse Ave., New Haven, CT 06511, USA}
\author{Tyler M. McCracken}
\affiliation{Ball Aerospace and Technologies Corporation, 1600 Commerce St, Boulder, CO 80301, USA}
\affiliation{Department of Astronomy, Yale University, 52 Hillhouse Ave., New Haven, CT 06511, USA}
\author[0000-0001-8483-4667]{David Sawyer}
\affiliation{Lowell Observatory, 1400 West Mars Hill Rd. Flagstaff, AZ 86001, USA}
\affiliation{Department of Astronomy, Yale University, 52 Hillhouse Ave., New Haven, CT 06511, USA}
\author[0000-0003-3047-5368]{Mikhail Sharov}
\affiliation{Department of Astronomy, Boston University, 725 Commonwealth Ave., Boston, MA 02215, USA}
\author[0000-0003-1001-0707]{René Tronsgaard}
\affiliation{DTU Space, National Space Institute, Technical University of Denmark, Elektrovej 328, DK-2800 Kgs. Lyngby, Denmark}


\author[0000-0002-4974-687X]{Andrew E. Szymkowiak}
\affiliation{Department of Astronomy, Yale University, 52 Hillhouse Ave., New Haven, CT 06511, USA}
\affiliation{Department of Physics, Yale University, 217 Prospect St, New Haven, CT 06511, USA}
\author[0000-0003-2221-0861]{Debra A. Fischer}
\affiliation{Department of Astronomy, Yale University, 52 Hillhouse Ave., New Haven, CT 06511, USA}

\newcommand{\unit}[1]{{\ensuremath{\,\mathrm{#1}}}}
\newcommand{\ms}{{\unit{m\,s^{-1}}}}
\newcommand{\cms}{{\unit{cm\,s^{-1}}}}
\newcommand{\um}{{\unit{\mu{} m}}}

\newcommand{\remark}[1]{{\color{red}#1}}

\newcommand{\annotate}[2]{\begin{tikzpicture}
    \node[anchor=south west,inner sep=0,align=center] (image) at (0,0) {
    #1
    };
    \begin{scope}[x={(image.south east)},y={(image.north west)}]
    #2
    \end{scope}
\end{tikzpicture}}

\newcommand{\replya}[1]{{\textbf{#1}}}

%% file: figures/pipeline.tex
\begin{tikzpicture}[y=2cm, x=2cm, inner sep=3pt, outer sep=0pt]

\node[draw](extflat) at (-5, 2.15){Extended Flat};
\node[draw](extdarks) at (-5, 1.89){Darks};

\node[draw, inner sep=1pt, fit=(extflat) (extdarks), label={}](extflats){};

\node[draw](masterflat) at (-2.75, 2.02){Master Flat};

\node[draw, blue, inner sep=4pt, fit=(extflats) (masterflat), label=above:\textcolor{blue}{Per Calibration Epoch}](epoch){};

\draw[->, red] (extflats) -- (masterflat) node [midway, above, label=below:{\scriptsize §3.1.3}]{\scriptsize Master Flat Reduction};

\node[draw](flat) at (-4.5, 1.40){Nightly Flats};
\node[draw](scienceraw) at (-4.5, 1.10){Science + Science Cals};

\node[draw, inner sep=1pt, fit=(flat) (scienceraw), label={}](2draw){};
\node[draw](dark) at (-4.5, 0.80){Darks};

\node[draw, blue, inner sep=4pt, fit=(dark) (2draw), label=below:\textcolor{blue}{Nightly Raw Exposures}](raw){};

\node[draw, red, rounded corners, label=above right:{\scriptsize \textcolor{red}{§3.1}}](reduction) at (-2.75, 1.25){Nightly Reduction};
\draw[->] (dark) -| (reduction) {};
\draw[->] (masterflat) -- (reduction) {};

\node[draw](redflat) at (-1.55, 1.55){Flats};
\node[draw](redscience) at (-1.55, 1.25){Science};
\node[draw](redcal) at (-1.55, 0.95){Cals};

\node[draw, inner sep=1pt, fit=(redcal) (redscience), label={}](2dred){};
\node[draw, blue, inner sep=4pt, fit=(redflat) (2dred), label=above:\textcolor{blue}{Reduced Exposures}](reduced){};

\draw[->, red] (2draw) -- (reduction) -- (reduced) {};

\node[draw, red, rounded corners, label=above right:{\scriptsize \textcolor{red}{§3.2}}](extraction) at (-0.25, 1.10){Optimal Extraction};
\draw[->] (redflat) -| (extraction) {};
\draw[->, red] (2dred) -- (extraction){};

\node[draw](thar) at (-0.25, 0){ThAr};
\node[draw](lfc) at (-0.25, -0.3){LFC};
\node[draw](science1d) at (-0.25, -0.8){Science};

\node[draw, inner sep=1pt, fit=(thar) (lfc), label=below:{Cals}](cals){};
\node[draw, blue, inner sep=4pt, fit=(cals) (science1d), label=below:\textcolor{blue}{Extracted 1D Spectra}](extracted){};
\draw[->, red] (extraction) -- (extracted){};

\node[draw, red, rounded corners, label=above right:{\scriptsize}, align=center](tharcal) at (-1.1, 0.475){ThAr Wavelength\\Calibration};
\node[draw, red, rounded corners, label=above right:{\scriptsize \textcolor{red}{§3.3}}, align=center](lfccal) at (-1.8, -0.3){LFC Wavelength\\Calibration};

\draw[->, red] (thar) -| (tharcal){};
\draw[->, red] (tharcal) -| (lfccal){};
\draw[->, red] (lfc) --  (lfccal){};

\node[draw](science) at (-3, -0.8){Science};
\node[draw](wave) at (-3, -1.1){Wavelength Solution};
\node[draw](barywave) at (-3, -1.4){Barycentric Wavelengths};
\node[draw](tell) at (-3, -1.7){Telluric Model};

\draw[->] (science1d) -- (science){};

\node[draw, double, forestgreen, inner sep=2pt, fit=(science) (wave) (barywave) (tell), label=above:\textcolor{forestgreen}{Finished FITS file}](fits){};

\draw[->, red] (lfccal) |- (wave){};

\node[draw, violet, align=center, label=above right:{\scriptsize \textcolor{violet}{Blackman+ (2019)}}](expm) at (-4.7, -0.6){Exposure Meter};
\node[draw, red, rounded corners, label=above right:{\scriptsize \textcolor{red}{§3.3}}](bc) at (-4.7, -1.1){Chromatic BC};

\draw[->] (expm) -- (bc){};
\draw[->, red] (wave) -- (bc){};
\draw[->, red] (bc) |- (barywave){};

\node[draw, violet, align=center, label=below left:{\scriptsize \textcolor{violet}{Leet+ (2019)}}](bstar) at (-4.5, -2.0){B star\\telluric database};

\draw[->] (fits) |- (bstar) {};
\draw[->] (bstar) |- (tell) {};

\node[draw, double, align=center, label=above right:{\scriptsize \textcolor{red}{§3.4.1}}](ccf) at (-0.9, -1.8){Linemask-based CCF;\\Coarse RV};
\node[draw, align=center, label=below left:{\scriptsize \textcolor{red}{§3.4.2}}](fwdmod) at (-0.9, -2.3){Chunk-by-chunk RV};

\node[draw, forestgreen, inner sep=12pt, fit=(ccf) (fwdmod), label=below:\textcolor{forestgreen}{Derived quantities}](derived){};

\draw[->, red](fits) -| (derived) {};
\draw[->, red] (derived) -- (1.5, -2.005) node [midway, above, label=below:{\scriptsize \textcolor{red}{§3.4.3}}]{Further Analysis};

\node[right](pub1) at (0.65, 0.25){\scriptsize\bf Legend};
\node[red, right](pub1) at (0.65, 0){\scriptsize Processes described in this work};
\node[violet, right](pub1) at (0.65, -.15){\scriptsize Processes described in other EXPRES papers};
\node[blue, right](pub1) at (0.65, -.50){\scriptsize Sets of exposures};
\node[forestgreen, right](pub1) at (0.65, -.65){\scriptsize Single exposures};

\end{tikzpicture}

%% file: vels.bbl
\begin{thebibliography}{}
\expandafter\ifx\csname natexlab\endcsname\relax\def\natexlab#1{#1}\fi
\providecommand{\url}[1]{\href{#1}{#1}}
\providecommand{\dodoi}[1]{doi:~\href{http://doi.org/#1}{\nolinkurl{#1}}}
\providecommand{\doeprint}[1]{\href{http://ascl.net/#1}{\nolinkurl{http://ascl.net/#1}}}
\providecommand{\doarXiv}[1]{\href{https://arxiv.org/abs/#1}{\nolinkurl{https://arxiv.org/abs/#1}}}

\bibitem[{{Anglada-Escud{\'e}} \& {Butler}(2012)}]{anglada_harps_2012}
{Anglada-Escud{\'e}}, G., \& {Butler}, R.~P. 2012, \apjs, 200, 15,
  \dodoi{10.1088/0067-0049/200/2/15}

\bibitem[{{Astropy Collaboration} {et~al.}(2013){Astropy Collaboration},
  {Robitaille}, {Tollerud}, {Greenfield}, {Droettboom}, {Bray}, {Aldcroft},
  {Davis}, {Ginsburg}, {Price-Whelan}, {Kerzendorf}, {Conley}, {Crighton},
  {Barbary}, {Muna}, {Ferguson}, {Grollier}, {Parikh}, {Nair}, {Unther},
  {Deil}, {Woillez}, {Conseil}, {Kramer}, {Turner}, {Singer}, {Fox}, {Weaver},
  {Zabalza}, {Edwards}, {Azalee Bostroem}, {Burke}, {Casey}, {Crawford},
  {Dencheva}, {Ely}, {Jenness}, {Labrie}, {Lim}, {Pierfederici}, {Pontzen},
  {Ptak}, {Refsdal}, {Servillat}, \& {Streicher}}]{astropy:2013}
{Astropy Collaboration}, {Robitaille}, T.~P., {Tollerud}, E.~J., {et~al.} 2013,
  \aap, 558, A33, \dodoi{10.1051/0004-6361/201322068}

\bibitem[{{Baranne} {et~al.}(1979){Baranne}, {Mayor}, \&
  {Poncet}}]{baranne_ccf_1979}
{Baranne}, A., {Mayor}, M., \& {Poncet}, J.~L. 1979, Vistas in Astronomy, 23,
  279, \dodoi{10.1016/0083-6656(79)90016-3}

\bibitem[{{Bedell} {et~al.}(2019){Bedell}, {Hogg}, {Foreman-Mackey}, {Montet},
  \& {Luger}}]{bedell_woble_2019}
{Bedell}, M., {Hogg}, D.~W., {Foreman-Mackey}, D., {Montet}, B.~T., \& {Luger},
  R. 2019, arXiv e-prints.
\newblock \doarXiv{1901.00503}

\bibitem[{{Blackman} {et~al.}(accepted){Blackman}, {Fischer}, {Jurgenson},
  {Sawyer}, {McCracken}, Szymkowiak, \& R.}]{blackman_performance_2020}
{Blackman}, R.~T., {Fischer}, D.~A., {Jurgenson}, C.~A., {et~al.} accepted, \aj

\bibitem[{{Blackman} {et~al.}(2019){Blackman}, {Ong}, \&
  {Fischer}}]{blackman_measured_2019}
{Blackman}, R.~T., {Ong}, J.~M.~J., \& {Fischer}, D.~A. 2019, \aj, 158, 40,
  \dodoi{10.3847/1538-3881/ab24c3}

\bibitem[{Blackman {et~al.}(2017)Blackman, Szymkowiak, Fischer, \&
  Jurgenson}]{blackman_accounting_2017}
Blackman, R.~T., Szymkowiak, A.~E., Fischer, D.~A., \& Jurgenson, C.~A. 2017,
  The Astrophysical Journal, 837, 18, \dodoi{10.3847/1538-4357/aa5ead}

\bibitem[{{Bolton} \& {Schlegel}(2010)}]{bolton_specperf_2010}
{Bolton}, A.~S., \& {Schlegel}, D.~J. 2010, \pasp, 122, 248,
  \dodoi{10.1086/651008}

\bibitem[{{Brahm} {et~al.}(2017){Brahm}, {Jord{\'a}n}, \&
  {Espinoza}}]{brahm_ceres_2017}
{Brahm}, R., {Jord{\'a}n}, A., \& {Espinoza}, N. 2017, \pasp, 129, 034002,
  \dodoi{10.1088/1538-3873/aa5455}

\bibitem[{{Butler} {et~al.}(1996){Butler}, {Marcy}, {Williams}, {McCarthy},
  {Dosanjh}, \& {Vogt}}]{butler_iodine_1996}
{Butler}, R.~P., {Marcy}, G.~W., {Williams}, E., {et~al.} 1996, \pasp, 108,
  500, \dodoi{10.1086/133755}

\bibitem[{{Ciddor}(1996)}]{ciddor_vactoair_1996}
{Ciddor}, P.~E. 1996, \ao, 35, 1566, \dodoi{10.1364/AO.35.001566}

\bibitem[{{Cornachione} {et~al.}(2019){Cornachione}, {Bolton}, {Eastman},
  {Wilson}, {Wang}, {Johnson}, {Sliski}, {McCrady}, {Wright}, {Plavchan},
  {Johnson}, {Horner}, \& {Wittenmyer}}]{cornachione_specperf_2019}
{Cornachione}, M.~A., {Bolton}, A.~S., {Eastman}, J.~D., {et~al.} 2019, \pasp,
  131, 124503, \dodoi{10.1088/1538-3873/ab4103}

\bibitem[{{Dumusque}(2018)}]{dumusque_measuring_2018}
{Dumusque}, X. 2018, \aap, 620, A47, \dodoi{10.1051/0004-6361/201833795}

\bibitem[{{Duncan} {et~al.}(1991){Duncan}, {Vaughan}, {Wilson}, {Preston},
  {Frazer}, {Lanning}, {Misch}, {Mueller}, {Soyumer}, {Woodard}, {Baliunas},
  {Noyes}, {Hartmann}, {Porter}, {Zwaan}, {Middelkoop}, {Rutten}, \&
  {Mihalas}}]{duncan_stellar_1991}
{Duncan}, D.~K., {Vaughan}, A.~H., {Wilson}, O.~C., {et~al.} 1991, \apjs, 76,
  383, \dodoi{10.1086/191572}

\bibitem[{{Fischer} {et~al.}(2016){Fischer}, {Anglada-Escude}, {Arriagada},
  {Baluev}, {Bean}, {Bouchy}, {Buchhave}, {Carroll}, {Chakraborty}, {Crepp},
  {Dawson}, {Diddams}, {Dumusque}, {Eastman}, {Endl}, {Figueira}, {Ford},
  {Foreman-Mackey}, {Fournier}, {F{\H{u}}r{\'e}sz}, {Gaudi}, {Gregory},
  {Grundahl}, {Hatzes}, {H{\'e}brard}, {Herrero}, {Hogg}, {Howard}, {Johnson},
  {Jorden}, {Jurgenson}, {Latham}, {Laughlin}, {Loredo}, {Lovis}, {Mahadevan},
  {McCracken}, {Pepe}, {Perez}, {Phillips}, {Plavchan}, {Prato}, {Quirrenbach},
  {Reiners}, {Robertson}, {Santos}, {Sawyer}, {Segransan}, {Sozzetti},
  {Steinmetz}, {Szentgyorgyi}, {Udry}, {Valenti}, {Wang}, {Wittenmyer}, \&
  {Wright}}]{fischer_eprv_2014}
{Fischer}, D.~A., {Anglada-Escude}, G., {Arriagada}, P., {et~al.} 2016, \pasp,
  128, 066001, \dodoi{10.1088/1538-3873/128/964/066001}

\bibitem[{{Freudling} {et~al.}(2013){Freudling}, {Romaniello}, {Bramich},
  {Ballester}, {Forchi}, {Garc{\'{\i}}a-Dabl{\'o}}, {Moehler}, \&
  {Neeser}}]{espresso_drs_2013}
{Freudling}, W., {Romaniello}, M., {Bramich}, D.~M., {et~al.} 2013, \aap, 559,
  A96, \dodoi{10.1051/0004-6361/201322494}

\bibitem[{{Gaia Collaboration}(2018)}]{gaiadr2}
{Gaia Collaboration}. 2018, VizieR Online Data Catalog, 1345

\bibitem[{{Horne}(1986)}]{horne_optimal_1986}
{Horne}, K. 1986, \pasp, 98, 609, \dodoi{10.1086/131801}

\bibitem[{{Isaacson} \& {Fischer}(2010)}]{isaacson_chromospheric_2010}
{Isaacson}, H., \& {Fischer}, D. 2010, \apj, 725, 875,
  \dodoi{10.1088/0004-637X/725/1/875}

\bibitem[{{Jurgenson} {et~al.}(2016){Jurgenson}, {Fischer}, {McCracken},
  {Sawyer}, {Szymkowiak}, {Davis}, {Muller}, \&
  {Santoro}}]{jurgenson_expres_2016}
{Jurgenson}, C., {Fischer}, D., {McCracken}, T., {et~al.} 2016, in \procspie,
  Vol. 9908, Ground-based and Airborne Instrumentation for Astronomy VI, 99086T

\bibitem[{{Landoni} {et~al.}(2014){Landoni}, {Riva}, {Pepe}, {Conconi},
  {Zerbi}, {Cabral}, {Cristiani}, \& {Megevand}}]{landoni_bc_2014}
{Landoni}, M., {Riva}, M., {Pepe}, F., {et~al.} 2014, Society of Photo-Optical
  Instrumentation Engineers (SPIE) Conference Series, Vol. 9147, {ESPRESSO
  front end exposure meter: a chromatic approach to radial velocity
  correction}, 91478K

\bibitem[{{Leet} {et~al.}(2019){Leet}, {Fischer}, \&
  {Valenti}}]{leet_tellurics_2019}
{Leet}, C., {Fischer}, D.~A., \& {Valenti}, J.~A. 2019, arXiv e-prints,
  arXiv:1903.08350.
\newblock \doarXiv{1903.08350}

\bibitem[{{Marcy} \& {Butler}(1992)}]{marcy_iodine_1992}
{Marcy}, G.~W., \& {Butler}, R.~P. 1992, \pasp, 104, 270,
  \dodoi{10.1086/132989}

\bibitem[{{Modigliani} {et~al.}(2019){Modigliani}, {Sownsowska}, \&
  {Lovis}}]{espresso_drs_2019}
{Modigliani}, A., {Sownsowska}, D., \& {Lovis}, C. 2019, ESPRESSO Pipeline User
  Manual, ESO

\bibitem[{Oliphant(2006--)}]{numpy}
Oliphant, T. 2006--, {NumPy}: A guide to {NumPy}, USA: Trelgol Publishing.
\newblock \url{http://www.numpy.org/}

\bibitem[{{Pepe} {et~al.}(2013){Pepe}, {Cristiani}, {Rebolo}, {Santos},
  {Dekker}, {M{\'e}gevand}, {Zerbi}, {Cabral}, {Molaro}, {Di Marcantonio},
  {Abreu}, {Affolter}, {Aliverti}, {Allende Prieto}, {Amate}, {Avila},
  {Baldini}, {Bristow}, {Broeg}, {Cirami}, {Coelho}, {Conconi}, {Coretti},
  {Cupani}, {D'Odorico}, {De Caprio}, {Delabre}, {Dorn}, {Figueira}, {Fragoso},
  {Galeotta}, {Genolet}, {Gomes}, {Gonz{\'a}lez Hern{\'a}ndez}, {Hughes},
  {Iwert}, {Kerber}, {Landoni}, {Lizon}, {Lovis}, {Maire}, {Mannetta},
  {Martins}, {Monteiro}, {Oliveira}, {Poretti}, {Rasilla}, {Riva}, {Santana
  Tschudi}, {Santos}, {Sosnowska}, {Sousa}, {Span{\`o}}, {Tenegi}, {Toso},
  {Vanzella}, {Viel}, \& {Zapatero Osorio}}]{pepe_espresso_2013}
{Pepe}, F., {Cristiani}, S., {Rebolo}, R., {et~al.} 2013, The Messenger, 153, 6

\bibitem[{{Piskunov} \& {Valenti}(2002)}]{piskunov_algorithms_2002}
{Piskunov}, N.~E., \& {Valenti}, J.~A. 2002, \aap, 385, 1095,
  \dodoi{10.1051/0004-6361:20020175}

\bibitem[{{Price-Whelan} {et~al.}(2018){Price-Whelan}, {Sip{\H{o}}cz},
  {G{\"u}nther}, {Lim}, {Crawford}, {Conseil}, {Shupe}, {Craig}, {Dencheva},
  {Ginsburg}, {VanderPlas}, {Bradley}, {P{\'e}rez-Su{\'a}rez}, {de Val-Borro},
  {Paper Contributors}, {Aldcroft}, {Cruz}, {Robitaille}, {Tollerud},
  {Coordination Committee}, {Ardelean}, {Babej}, {Bach}, {Bachetti}, {Bakanov},
  {Bamford}, {Barentsen}, {Barmby}, {Baumbach}, {Berry}, {Biscani}, {Boquien},
  {Bostroem}, {Bouma}, {Brammer}, {Bray}, {Breytenbach}, {Buddelmeijer},
  {Burke}, {Calderone}, {Cano Rodr{\'\i}guez}, {Cara}, {Cardoso}, {Cheedella},
  {Copin}, {Corrales}, {Crichton}, {D{\textquoteright}Avella}, {Deil},
  {Depagne}, {Dietrich}, {Donath}, {Droettboom}, {Earl}, {Erben}, {Fabbro},
  {Ferreira}, {Finethy}, {Fox}, {Garrison}, {Gibbons}, {Goldstein}, {Gommers},
  {Greco}, {Greenfield}, {Groener}, {Grollier}, {Hagen}, {Hirst}, {Homeier},
  {Horton}, {Hosseinzadeh}, {Hu}, {Hunkeler}, {Ivezi{\'c}}, {Jain}, {Jenness},
  {Kanarek}, {Kendrew}, {Kern}, {Kerzendorf}, {Khvalko}, {King}, {Kirkby},
  {Kulkarni}, {Kumar}, {Lee}, {Lenz}, {Littlefair}, {Ma}, {Macleod},
  {Mastropietro}, {McCully}, {Montagnac}, {Morris}, {Mueller}, {Mumford},
  {Muna}, {Murphy}, {Nelson}, {Nguyen}, {Ninan}, {N{\"o}the}, {Ogaz}, {Oh},
  {Parejko}, {Parley}, {Pascual}, {Patil}, {Patil}, {Plunkett}, {Prochaska},
  {Rastogi}, {Reddy Janga}, {Sabater}, {Sakurikar}, {Seifert}, {Sherbert},
  {Sherwood-Taylor}, {Shih}, {Sick}, {Silbiger}, {Singanamalla}, {Singer},
  {Sladen}, {Sooley}, {Sornarajah}, {Streicher}, {Teuben}, {Thomas},
  {Tremblay}, {Turner}, {Terr{\'o}n}, {van Kerkwijk}, {de la Vega}, {Watkins},
  {Weaver}, {Whitmore}, {Woillez}, {Zabalza}, \& {Contributors}}]{astropy:2018}
{Price-Whelan}, A.~M., {Sip{\H{o}}cz}, B.~M., {G{\"u}nther}, H.~M., {et~al.}
  2018, \aj, 156, 123, \dodoi{10.3847/1538-3881/aabc4f}

\bibitem[{{Probst} {et~al.}(2016){Probst}, {Lo Curto}, {{\'A}vila},
  {Brucalassi}, {Canto Martins}, {de Castro Le{\~a}o}, {Esposito},
  {Gonz{\'a}lez Hern{\'a}ndez}, {Grupp}, {H{\"a}nsch}, {Holzwarth},
  {Kellermann}, {Kerber}, {Mandel}, {Manescau}, {Pasquini}, {Pozna}, {Rebolo},
  {Renan de Medeiros}, {Stark}, {Steinmetz}, {Su{\'a}rez Mascare{\~n}o},
  {Udem}, {Urrutia}, \& {Wu}}]{probst_relative_2016}
{Probst}, R.~A., {Lo Curto}, G., {{\'A}vila}, G., {et~al.} 2016, in \procspie,
  Vol. 9908, Ground-based and Airborne Instrumentation for Astronomy VI, 990864

\bibitem[{{Schwab} {et~al.}(2016){Schwab}, {Rakich}, {Gong}, {Mahadevan},
  {Halverson}, {Roy}, {Terrien}, {Robertson}, {Hearty}, {Levi}, {Monson},
  {Wright}, {McElwain}, {Bender}, {Blake}, {St{\"u}rmer}, {Gurevich},
  {Chakraborty}, \& {Ramsey}}]{neid}
{Schwab}, C., {Rakich}, A., {Gong}, Q., {et~al.} 2016, in \procspie, Vol. 9908,
  Ground-based and Airborne Instrumentation for Astronomy VI, 99087H

\bibitem[{{Steinmetz} {et~al.}(2008){Steinmetz}, {Wilken}, {Araujo-Hauck},
  {Holzwarth}, {H{\"a}nsch}, {Pasquini}, {Manescau}, {D'Odorico}, {Murphy},
  {Kentischer}, {Schmidt}, \& {Udem}}]{steinmetz_laser_2008}
{Steinmetz}, T., {Wilken}, T., {Araujo-Hauck}, C., {et~al.} 2008, Science, 321,
  1335, \dodoi{10.1126/science.1161030}

\bibitem[{{Szymkowiak} {et~al.}(in preparation){Szymkowiak}, {Fischer},
  {et~al.}}]{szymkowiak_lfc_2019}
{Szymkowiak}, A., {Fischer}, D.~A., {et~al.} in preparation

\bibitem[{{Tal-Or} {et~al.}(2019){Tal-Or}, {Trifonov}, {Zucker}, {Mazeh}, \&
  {Zechmeister}}]{talor_correcting_2019}
{Tal-Or}, L., {Trifonov}, T., {Zucker}, S., {Mazeh}, T., \& {Zechmeister}, M.
  2019, \mnras, 484, L8, \dodoi{10.1093/mnrasl/sly227}

\bibitem[{{Trifonov} {et~al.}(2020){Trifonov}, {Tal-Or}, {Zechmeister},
  {Kaminski}, {Zucker}, \& {Mazeh}}]{trifonov_harps_2020}
{Trifonov}, T., {Tal-Or}, L., {Zechmeister}, M., {et~al.} 2020, arXiv e-prints,
  arXiv:2001.05942.
\newblock \doarXiv{2001.05942}

\bibitem[{{Tronsgaard} {et~al.}(2019){Tronsgaard}, {Buchhave}, {Wright},
  {Eastman}, \& {Blackman}}]{tronsgaard_bc_2019}
{Tronsgaard}, R., {Buchhave}, L.~A., {Wright}, J.~T., {Eastman}, J.~D., \&
  {Blackman}, R.~T. 2019, \mnras, 489, 2395, \dodoi{10.1093/mnras/stz2181}

\bibitem[{{van der Walt} {et~al.}(2011){van der Walt}, {Colbert}, \&
  {Varoquaux}}]{numpy2}
{van der Walt}, S., {Colbert}, S.~C., \& {Varoquaux}, G. 2011, Computing in
  Science Engineering, 13, 22, \dodoi{10.1109/MCSE.2011.37}

\bibitem[{{Virtanen} {et~al.}(2020){Virtanen}, {Gommers}, {Oliphant},
  {Haberland}, {Reddy}, {Cournapeau}, {Burovski}, {Peterson}, {Weckesser},
  {Bright}, {van der Walt}, {Brett}, {Wilson}, {Jarrod Millman}, {Mayorov},
  {Nelson}, {Jones}, {Kern}, {Larson}, {Carey}, {Polat}, {Feng}, {Moore}, {Vand
  erPlas}, {Laxalde}, {Perktold}, {Cimrman}, {Henriksen}, {Quintero}, {Harris},
  {Archibald}, {Ribeiro}, {Pedregosa}, {van Mulbregt}, \&
  {Contributors}}]{scipy}
{Virtanen}, P., {Gommers}, R., {Oliphant}, T.~E., {et~al.} 2020, Nature
  Methods, 17, 261, \dodoi{https://doi.org/10.1038/s41592-019-0686-2}

\bibitem[{{Wang} \& {Ford}(2011)}]{wang_eccentricity_2011}
{Wang}, J., \& {Ford}, E.~B. 2011, \mnras, 418, 1822,
  \dodoi{10.1111/j.1365-2966.2011.19600.x}

\bibitem[{{Wilson} {et~al.}(2019){Wilson}, {Eastman}, {Cornachione}, {Wang},
  {Johnson}, {Sliski}, {Schap}, {Morton}, {Johnson}, {McCrady}, {Wright},
  {Wittenmyer}, {Plavchan}, {Blake}, {Swift}, {Bottom}, {Baker}, {Barnes},
  {Berlind}, {Blackhurst}, {Beatty}, {Bolton}, {Cale}, {Calkins}, {Col{\'o}n},
  {de Vera}, {Esquerdo}, {Falco}, {Fortin}, {Garcia-Mejia}, {Geneser},
  {Gibson}, {Grell}, {Groner}, {Halverson}, {Hamlin}, {Henderson}, {Horner},
  {Houghton}, {Janssens}, {Jonas}, {Jones}, {Kirby}, {Lawrence}, {Luebbers},
  {Muirhead}, {Myles}, {Nava}, {Rivera-Garc{\'\i}a}, {Reed}, {Relles},
  {Riddle}, {Robinson}, {Chaput de Saintonge}, \& {Sergi}}]{wilson2019}
{Wilson}, M.~L., {Eastman}, J.~D., {Cornachione}, M.~A., {et~al.} 2019, \pasp,
  131, 115001, \dodoi{10.1088/1538-3873/ab33c5}

\bibitem[{{Winn} \& {Fabrycky}(2015)}]{winn_occurrence_2015}
{Winn}, J.~N., \& {Fabrycky}, D.~C. 2015, \araa, 53, 409,
  \dodoi{10.1146/annurev-astro-082214-122246}

\bibitem[{Wright \& Eastman(2014)}]{wright_barycorr_2014}
Wright, J.~T., \& Eastman, J.~D. 2014, Publications of the Astronomical Society
  of the Pacific, 126, 838–852, \dodoi{10.1086/678541}

\bibitem[{{Xu} {et~al.}(2019){Xu}, {Cisewski-Kehe}, {Davis}, {Fischer}, \&
  {Brewer}}]{xu_alpha_hull_2019}
{Xu}, X., {Cisewski-Kehe}, J., {Davis}, A.~B., {Fischer}, D.~A., \& {Brewer},
  J.~M. 2019, \aj, 157, 243, \dodoi{10.3847/1538-3881/ab1b47}

\bibitem[{{Zechmeister} {et~al.}(2014){Zechmeister}, {Anglada-Escud{\'e}}, \&
  {Reiners}}]{zechmeister_flat_2014}
{Zechmeister}, M., {Anglada-Escud{\'e}}, G., \& {Reiners}, A. 2014, \aap, 561,
  A59, \dodoi{10.1051/0004-6361/201322746}

\end{thebibliography}
